\documentclass[aps,pra,nopacs,oneside,floatfix,a4,superscriptaddress,twocolumn,nofootinbib]{revtex4-1}

\usepackage{graphicx}
\usepackage{dsfont}
\newtheorem{thm}{Theorem}
\newtheorem{lem}[thm]{Lemma}

\usepackage{amsfonts,amsmath,amssymb}
\usepackage{hyperref}
\hypersetup{colorlinks=false,pdfborder={0 0 0}}
\usepackage[utf8]{inputenc}
\usepackage{color,verbatim}
\usepackage{times}
\usepackage{enumerate}

\newcommand{\ba}{\begin{eqnarray}}
\newcommand{\be}{\begin{equation}}
\newcommand{\ee}{\end{equation}}
\newcommand{\beq}{\begin{equation}}
\newcommand{\eeq}{  \end{equation}}
\newcommand{\bea}{\begin{eqnarray}}
\newcommand{\eea}{  \end{eqnarray}}
\newcommand{\g}{\textrm{gen}}
\newcommand{\cc}{\textrm{cl}}
\newcommand{\rr}{\textrm{r}}

\newcommand{\ea}{\end{eqnarray}}
\newcommand{\ban}{\begin{eqnarray*}}
\newcommand{\ean}{\end{eqnarray*}}

\newcommand{\tr}{\operatorname{tr}}

\newcommand{\ket}[1]{\vert #1 \rangle}
\newcommand{\bra}[1]{\langle #1 \vert}

\newcommand{\Tel}{\{\sigma_{a|\omega_x}\}_{a,x}}
\newcommand{\tel}{\sigma_{a|\omega_x}}

\newcommand{\eg}{{\it{e.g.}~}}
\newcommand{\ie}{{\it{i.e.}~}}

\newcommand{\rA}{\mathrm{A}}

\newcommand{\rB}{\mathrm{B}}
\newcommand{\rV}{\mathrm{V}}
\newcommand{\rC}{\mathrm{C}}

\newcommand{\BSA}{\mathrm{BSA}}

\definecolor{ivan}{rgb}{0.7,0,0.7}
 
\begin{document}

\preprint{APS/123-QED}

\title{Estimating entanglement in teleportation experiments}

\author{Ivan \v{S}upi\'{c}}
\affiliation{ICFO-Institut de Ciencies Fotoniques,  The Barcelona Institute of Science and Technology,  08860 Castelldefels (Barcelona),  Spain}

\author{Paul Skrzypczyk}
\affiliation{H. H. Wills Physics Laboratory, University of Bristol, Tyndall Avenue, Bristol, BS8 1TL, United Kingdom}%

\author{Daniel Cavalcanti}
\affiliation{ICFO-Institut de Ciencies Fotoniques,  The Barcelona Institute of Science and Technology,  08860 Castelldefels (Barcelona),  Spain}

\date{\today}

\begin{abstract}
Quantum state teleportation is a protocol where a shared entangled state is used as a quantum channel to transmit quantum information between distinct locations. Here we consider the task of estimating entanglement in teleportation experiments. We show that the data accessible in a teleportation experiment allows to put a lower bound on some entanglement measures, such as entanglement negativity and robustness. Furthermore, we show cases in which the lower bounds are tight. The introduced lower bounds can also be interpreted as quantifiers of the nonclassicality of a teleportation experiment. Thus, our findings provide a quantitative relation  between teleportation and entanglement. 
\end{abstract}

\maketitle


\section{Introduction}

The seminal work by Bennett et al \cite{teleportation} from 1993 demonstrated the possibility to faithfully transfer the quantum state of a system to a spatially distant one, without having to physically send it. Named quantum teleportation, this protocol made a huge impact on the development of quantum information processing, being a building block for more advanced protocols such as cryptographic tasks \cite{Gisin}, quantum repeaters \cite{repeaters}, quantum computing \cite{Gottesman, Raussendorf} and many others. \\

Ideally, in order to realise teleportation two parties, Alice and Bob, need to share a pair of particles in a maximally entangled state $\ket{\Phi^+} = \sum_{i=0}^{d-1}\ket{ii}/\sqrt{d}$, where $d$ is the local Hilbert space dimension of the system. Then, Alice applies a joint Bell state measurement (a measurement where all measurement operators are maximally entangled) on a third system in state $\ket{\omega}$ and her share of the maximally entangled state and communicates the result to Bob. Bob, upon receiving the message from Alice, applies a unitary operation on his system, which ends up in the desired state $\ket{\omega}$. A very important feature of quantum teleportation is that Alice does not need to know the state $\ket{\omega}$.

In realistic conditions it is impossible to achieve perfect quantum teleportation. There has been a lot of effort in describing imperfect teleportation as well as the role of generic entangled states in the protocol. In a teleportation experiment where a set of $N_x$ states $\{\omega_x\}_{x=1}^{N_x}$ -- which need not necessarily be pure states -- is teleported, the most common benchmark between classical and quantum teleportation is the average fidelity of teleportation \cite{Sandu}
\ba\label{e:fidelity}
\overline{F}_{\tel}=\frac{1}{N_x}\sum_{a,x} p(a|\omega_x)F(U_a \rho^{\rB}_{a|\omega_x} U_a^{\dagger}, \omega_x)
\ea
where $F(\rho,\sigma) = \|\sqrt{\rho}\sqrt{\sigma}\|_1$ is the fidelity, and 
\be\label{Bobassemblage}
\rho_{a|\omega_x}^\rB=\frac{\tr_{\rV\rA} [(M_{a}^{\rV\rA}\otimes \openone^\rB)(\omega_x^\rV\otimes\rho^{\rA\rB})]}{p(a|\omega_x)}
\ee 
are the states Bob obtains, conditioned on the input state $\omega_x$ and Alice's measurement output $a$, while $p(a|\omega_x)$ is the probability for Alice to obtain the outcome $a$ when the input state is $\omega_x$. The teleportation process is considered to be quantum (\ie non-classical) if the fidelity of teleportation is higher than the fidelity that could be obtained using solely classical resources (\ie no entanglement pre-shared between Alice and Bob). Based on this figure of merit, not all entangled states are useful for achieving non-classical teleportation \cite{Sandu,Horodecki99}, among them the bound entangled states \cite{bound entanglement}.\\

Notice that the average fidelity of teleportation is a coarse grained measure, reducing all the information available in a teleportation experiment to a single number. Thus, it could happen that, even though an entangled state could not achieve an average fidelity higher that a separable state, a deeper analysis of the relation between input and output states, summarized by the states \eqref{Bobassemblage} and observed statistics $p(a|\omega_x)$, could lead to a better assessment on the non-classical nature of the teleportation process. Motivated by this, in our recent work \cite{CSS} we proved that the information available in a teleportation experiment allows to prove that every entangled state can be used to demonstrate a nonclassical teleportation process, if a suitable set of input states and measurement are chosen. The method was moreover used to experimentally demonstrate non-classical teleportation stemming from a state unable to outperform separable states in terms of the average fidelity of teleportation \cite{Rome}.

Our main goal is to go beyond detection of nonclassicality, and to show that the teleportation data  can also be used to estimate the amount of entanglement of the state shared between Alice and Bob. This can be done in two ways: by considering the full teleportation data (the correlations between input and output states) or through the violation of teleportation witnesses (linear functions of teleportation data).\\

Let us recall the formalism for describing quantum state teleportation introduced in \cite{CSS}. Quantum teleportation manifests in the nonlocal correlations between Alice's joint measurement outputs and states prepared for Bob. A teleportation experiment is non-classical if it excludes a ``local-hidden-channel model", which would, in a classical way, correlate Alice's outputs with Bob's reduced states. In order to see exactly the form of such classical teleportation channels let us see what data Alice and Bob can observe if they share a separable state. The set of reduced states of Bob forms a teleportation assemblage (teleportage in \cite{HobanSainz}). In a teleportation experiment, Bob's (unnormalised) state, given Alice's input state is $\omega_x^V$, is given by
\ba \label{e:unnstates}
\sigma_{a|\omega_x}^\rB &=& \tr_{\rV\rA} [(M_{a}^{\rV\rA}\otimes \openone^\rB)(\omega_x^\rV\otimes\rho^{\rA\rB})] \nonumber \\
&=& \tr_{\rV} [\tilde{M}_{a}^{\rV\rB}(\omega_x^\rV\otimes\openone^{\rB})],
\ea
where $M_{a}^{\rV\rA}$ are the operators describing the measurement happening inside Alice's box, and
\be \label{choperators}
\tilde{M}_{a}^{\rV\rB}=\tr_\rA[(M_{a}^{\rV\rA}\otimes \openone^\rB)(\openone^\rV\otimes\rho^{\rA\rB})].
\ee
Notice that the normalization of $\sigma_{a|\omega_x}^\rB$ gives the probabilities of Alice's outcomes $p(a|\omega_x)$.

If $\rho^{\rA\rB}$ is separable, \ie $\rho^{\rA\rB} = \sum_\lambda p_\lambda \rho^\rA_\lambda\otimes\rho^\rB_\lambda$, then \cite{CSS}
\be \label{LocModChOP}
\tilde{M}_{a}^{\rV\rB} = \sum_\lambda p_\lambda M_{a|\lambda}^{\rV}\otimes\rho^\rB_\lambda,
\ee
where
\ba
M_{a|\lambda}^{\rV}= \tr_\rA[ M_{a}^{\rV\rA} (\openone^\rV\otimes\rho^\rA_\lambda)].
\ea
We call this case a local-hidden-channel model for the teleportation experiment, since it can be understood in the following way: at each round of the experiment a classical variable $\lambda$ is sent to Alice and Bob. Upon reading the value of $\lambda$ Alice's device output $a$ with probability $p(a|\omega_x,\lambda)=\tr [M_{a|\lambda}^\rV\omega_x^\rV]$ and Bob's device generates the state $\rho_\lambda$. 

Given an observed teleportation data, \ie$\{\sigma_{a|\omega_x}^B\}$, such classical model can be tested by solving the following optimization problem:
\begin{align}\label{eq:feasibility}\begin{split}
\text{given}& \quad\{\sigma_{a|\omega_x}^\rB\}_{a,x}, \{\omega_x^\rV\}_x \\
\text{find}& \quad \{\tilde{M}_a^{\rV\rB}\}_a  \\
\text{s.t.}&\quad  \sigma_{a|\omega_x}^\rB=\tr [\tilde{M}_a^{\rV\rB} (\omega_x^\rV\otimes\openone^\rB)] \quad\forall a,x, \\
&\quad \tilde{M}_a^{\rV\rB} \in \mathcal{S} \quad \forall a, \end{split}
\end{align}
where $\mathcal{S}$ denotes the set of separable operators, \ie operators of the form $\sum_\lambda \tau_\lambda \otimes \chi_\lambda$, with $\tau_\lambda \geq 0$ and $\chi_\lambda \geq 0$ for all $\lambda$. This set cannot be characterised in a simple way, but it relaxations to the set of operators with positive partial transpose (PPT) or the set of $k$-shareable operators \cite{Doherty} can be characterised through linear SDP constraints \cite{SDP}. \\

In Ref. \cite{CSS} we proved that with a suitable set of input states and a suitable measurement for Alice, every entangled state leads to a non-classical teleportation data, in the sense of not having the model \eqref{e:unnstates} and \eqref{LocModChOP}. \\

To conclude the introduction, let us note that there are two  elements contributing to the constraint that the channel operators (\ref{choperators}) are separable: the shared state $\rho^{\rA\rB}$ is separable; or Alice's measurement operators $M_a^{\rV\rA}$ are not entangling operators. Indeed, taking  $M_a^{\rV\rA} = \sum_{\lambda} \tau_{\lambda,a}^\rV \otimes \chi_{\lambda,a}^\rA$ makes the channel operators separable:
\ba\label{savnik}
\tilde{M}_a^{\rV\rB} &=& \tr_\rA[(\sum_{\lambda}\tau_{\lambda,a}^\rV \otimes \chi_{\lambda,a}^\rA\otimes \mathds{1}^\rB)(\mathds{1}^\rV\otimes \rho^{\rA\rB})]\nonumber \\
&=& \sum_{\lambda} \tau_{\lambda,a}^\rV \otimes \pi_{\lambda,a}^\rB,
\ea
where $\pi^\rB_{\lambda,a} = \tr[(\chi_{\lambda,a}^\rA\otimes \mathds{1}^\rB)\rho^{\rA\rB}]$. With this in mind, infeasibility of the problem (\ref{eq:feasibility}) certifies two things simultaneously:
\begin{itemize}
\item[$\star$]{the state $\rho^{\rA\rB}$ is entangled},
\item[$\star$]{the measurements $M_a^{\rV\rA}$ are entangling.}
\end{itemize}

In the rest of this paper we show that entanglement not only leads to qualitatively different teleportation experiments than separable states, but it can also be quantified from a teleportation experiment. In Section \ref{sec:neg} we discuss possibilities to quantify entanglement negativity from a teleportation experiment. In Section \ref{rob} we put a lower bound on robustness of entanglement and define various robustness-based teleportation quantifiers. Finally, in Section \ref{weight} we introduce teleportation weight and relate it to the best separable approximation of an entangled state.

\section{Estimating entanglement negativity from a teleportation experiment}\label{sec:neg}


Negativity of entanglement \cite{VW} is a widely used entanglement measure, largely due to the fact that it can be computed efficiently. In the original paper introducing entanglement negativity \cite{VW} the authors examined relation between 'teleportation capacity' of a quantum state and its negativity. More precisely, the authors proved that the average fidelity of teleoportation puts a lower bound on the entanglement negativity of the shared state. Since we introduced a way to characterise teleportation experiment beyond average fidelity, we expect that our method can give some further insight into the role of entanglement negativity in teleportation. In this section we prove that, indeed, by using all the accessible information in a teleportation experiment one can place lower bound on the entanglement negativity of the shared state $\rho^{\rA\rB}$. \\

\subsection{Estimating entanglement negativity from teleportation data}

The entanglement negativity of a state $\rho^{\rA\rB}$ can be expressed in the following way
\begin{align}\label{neg}\begin{split}
\mathcal{N}(\rho_{\rA\rB}) &= \min_{\rho_+,\rho_-} \quad \textrm{tr}(\rho_-) \\ 
& \textrm{s.t.} \quad \rho^{\rA\rB} = \rho_+ - \rho_- \\ 
&\quad \quad {\rho_{\pm}}^{T_\rA} \geq 0. \end{split}
\end{align}
Let us see how this optimization problem can be rephrased in terms of the available teleportation data, \ie the teleportation assemblage $\{\sigma_{a\vert \omega_x}\}_{a,x}$. Analogously to \eqref{e:unnstates} we can introduce the auxiliary teleportation assemblages
\begin{equation*}
\sigma_{a\vert \omega_x}^{\pm} = \textrm{tr}_{\rV\rA}\left[\left(M_a^{\rV\rA}\otimes \mathds{1}^\rB\right)\left(\omega_x^\rV\otimes \rho_{\pm}^{\rA\rB}\right)\right].
\end{equation*}
With this notation the first constraint from \eqref{neg} can be written as 
\begin{equation*}
\sigma_{a\vert \omega_x} = \sigma_{a\vert \omega_x}^{+} - \sigma_{a\vert \omega_x}^{-}
\end{equation*}
The objective function is easily identified as
\begin{equation}\label{objfunc}
\sum_a \textrm{tr}\left[ \sigma_{a\vert \omega_x}^-\right] = \textrm{tr}\left[\omega_x^\rV \otimes \rho_-^{\rA\rB}\right] = \textrm{tr}\left[\rho_-^{\rA\rB}\right]
\end{equation}
Finally we have to characterize the effective measurements $M_{a,\pm}^{*\rV\rB}$ which can arise from a PPT state. From (\ref{choperators}) and the identity 
\begin{equation}\label{MaxEntIdn}
\tr_{\rB}\left[\left(\mathds{1}^{\rA}\otimes M^{\rB\rC} \right)\left( {\Phi^+}^{\rA\rB}\otimes\mathds{1}^{\rC}\right)\right] = \frac{1}{d}\left(M^{\rA\rC}\right)^{T_{\rA}}
\end{equation}
we obtain
\begin{multline}\label{effmeasPPT}
d\left(M_a^{*\rV\rB}\right)^T \\= \textrm{tr}_{\rV_1\rA}\left[\left(\mathds{1}^{\rV}\otimes M_a^{\rV_1\rA}\otimes \mathds{1}^\rB\right)\left({\Phi^+}^{\rV\rV_1}\otimes {\rho^{\rA\rB}}^{T_\rB}\right)\right].
\end{multline}
If the state $\rho^{\rA\rB}$ is PPT the left hand side of the last equation represents an unnormalized quantum state, which means that the operators $\left({M_a^*}^{\rV\rB}\right)^T$ and hence $\left({M_a^*}^{\rV\rB}\right)$ are also positive, which justifies the last constraint from (\ref{negtel}). Now we are ready to construct a semidefinite program whose solution represents a lower bound on the negativity of the shared state $\rho^{\rA\rB}$:
\begin{align}\label{negtel}
\begin{split}
\min_{\{M_{a,\pm}^{*}\}_a,\rho_\pm} &\quad \sum_a \textrm{tr}\left[ \sigma_{a\vert \omega_x}^-\right] \\ 
\textrm{s.t.} &\quad \sigma_{a\vert \omega_x} = \sigma_{a\vert \omega_x}^{+} - \sigma_{a\vert \omega_x}^{-} \\ 
&\quad  \sigma_{a\vert \omega_x}^{\pm} = \textrm{tr}_\rV\left[{M_{a,\pm}^*}^{\rV\rB}\left(\omega_x^\rV\otimes \mathds{1}^\rB\right)\right] \\
&\quad \sum_a {M_{a,\pm}^*}^{\rV\rB} = \openone^\rV \otimes \rho_\pm^\rB, \\
&\quad  M_{a,\pm}^{*\rV\rB} \geq 0.
\end{split}
\end{align}
The next step is to prove that the solution of (\ref{negtel}) lower bounds the negativity of entanglement of the state $\rho^{\rA\rB}$, given by the solution of (\ref{neg}). First let us note that in the case the set of input states is tomographically complete and Alice applies the Bell state measurement, the solutions to (\ref{neg}) and (\ref{negtel}) coincide. To see that, let us rewrite the optimization problem (\ref{negtel}) for $a=0$ given that $M_0^{\rV\rA} = {\Phi^+}^{\rV\rA}$:
\begin{align}\label{negtelBSM}
\begin{split}
\min_{{\rho}_{\pm}} &\quad \textrm{tr}[\rho_-] \\ 
\textrm{s.t.} &\quad \textrm{tr}_{\rV\rA}\left[\left({\Phi^+}^{\rV\rA}\otimes \mathds{1}^\rB\right)\left(\omega_x^\rV \otimes \rho^{\rA\rB}\right)\right] \\  &\quad = \textrm{tr}_{\rV\rA}\left[\left({\Phi^+}^{\rV\rA}\otimes \mathds{1}^\rB\right)\left(\omega_x^\rV \otimes\left( \rho_+^{\rA\rB}-\rho_-^{\rA\rB}\right)\right)\right],\\ 
&\quad  \sigma_{0\vert \omega_x}^{\pm} = \textrm{tr}_\rV\left[M_{0,\pm}^{*\rV\rB}\left(\omega_x^\rV\otimes \mathds{1}^\rB\right)\right], \\ 
&\quad  M_{0,\pm}^{*\rV\rB} = \textrm{tr}_{\rA}\left[\left({\Phi^+}^{\rV\rA}\otimes \mathds{1}^\rB\right)\left(\mathds{1}^\rV\otimes \rho_{\pm}^{\rA\rB}\right)\right]  \\ &\hspace{5cm}= \frac{1}{d}\rho_{\pm}^{T_\rB} \geq 0.
\end{split}
\end{align}
The first constraint in case of a tomographically complete set of inputs is satisfied if and only if
\begin{equation*}
\rho_{\rA\rB} = \rho_+ - \rho_-,
\end{equation*}
which finally reduces (\ref{negtelBSM}) to (\ref{neg}). For the other values of $a$ the constraints from \eqref{negtelBSM} are automatically satisfied. The first constraint can be rewritten as 
\begin{equation*}
\begin{split}
&\quad \textrm{tr}_{\rV\rA}\left[\left(U_a^\rV{\Phi^+}^{\rV\rA}{U_a^{\dagger}}^\rV\otimes \mathds{1}^\rB\right)\left(\omega_x^\rV \otimes \rho^{\rA\rB}\right)\right] = \\  &\quad = \textrm{tr}_{\rV\rA}\left[\left(U_a^\rV{\Phi^+}^{\rV\rA}{U_a^{\dagger}}^\rV\otimes \mathds{1}^\rB\right)\left(\omega_x^\rV \otimes\left( \rho_+^{\rA\rB}-\rho_-^{\rA\rB}\right)\right)\right]
\end{split}
\end{equation*}
which is equivalent to 
\begin{multline*}
\textrm{tr}_{\rV\rA}\left[\left({\Phi^+}^{\rV\rA}\otimes \mathds{1}^\rB\right)\left({U_a^{\dagger}}^\rV\omega_x^\rV U_a^\rV \otimes \rho^{\rA\rB}\right)\right]  \\  = \textrm{tr}_{\rV\rA}\left[\left({\Phi^+}^{\rV\rA}\otimes \mathds{1}^\rB\right)\left({U_a^{\dagger}}^\rV\omega_x^\rV U_a^\rV \otimes\left( \rho_+^{\rA\rB}-\rho_-^{\rA\rB}\right)\right)\right].
\end{multline*}
If the set $\{\omega_x\}_x$ is tomographically complete, so is $\{U_a^{\dagger}\omega_x U_a\}_{x}$, and thus the last statement is equivalent to $\rho_{\rA\rB} = \rho_+ - \rho_-$. Similarly the last constraint from \eqref{negtelBSM} reduces to $U_a\rho_{\pm}^{T_\rB}U_a^{\dagger} \geq 0$, which is satisfied if $\rho_{\pm}^{T_\rB} \geq 0$. Thus, we see that when Alice applies the full Bell state measurement and has access to a tomographically complete set of input states, the optimization problems \eqref{negtel} and \eqref{neg} are equivalent.\\

In the general case, note that the states $\rho'_{\pm}$ leading to the optimal solution of \eqref{neg} by forming $\sigma_{a|\omega_x}^{\pm} = \tr_{\rV\rA}[(M_a^{\rV\rA}\otimes \mathds{1}^\rB)(\omega_x^\rV\otimes \rho_{\pm}^{\rA\rB})]$  with arbitrary measurements $\{M_a^{\rV\rA}\}$ and input states $\{\omega_x\}_x$ satisfy all the constraints of \eqref{negtel}. The equivalence between the objective functions follows from   \eqref{objfunc} and the last constraint is satisfied due to \eqref{effmeasPPT}. This means that the solution to \eqref{negtel} cannot be larger than $\mathcal{N}(\rho^{\rA\rB})$, \ie it places a lower bound on the entanglement negativity of $\rho^{\rA\rB}$.

\subsection{Estimating entanglement negativaty from violations of nonclassical teleportation witnesses}

In \citep{CSS} we introduced an SDP optimization problem which determines if observed teleportation data can be reproduced with classical teleportation channels. The dual form of the introduced SDP gives a nonclassical teleportation witness, i.e. an operator which is positive whenever evaluated on Bob's states obtained from a classical teleportation experiment, but can take a negative value when evaluated on states resulting from a nonclassical teleportation. Violation of nonclassical teleportation witness was used to experimentally certify nonclassicality of a teleportation protocol in \cite{Rome}. In this subsection we show that teleportation witnesses, besides certifying nonclassical teleportation, can be used to put a lower bound on entanglement negativity of the shared state. The use of teleportation witnesses may be favourable since knowing the full teleportation assemblage requires performing quantum state tomography which is often a costly task. Contrarily, one can obtain the violation of a teleportation witness without performing the full tomography of the teleported states. \\

Let us observe that the average teleportation fidelity represents a particular type of teleportation witness, and remind again that the authors of \cite{VW} proved that it lower bounds entanglement negativity. Here we provide an SDP which allows to estimate entanglement negativity from violation of an arbitrary teleportation witness $F_{a|\omega_x}$. Assuming that the observed violation of the witness is $w$, the following SDP provides a lower bound on the negativity of the shared state
\begin{align}
&\mathcal{N}(\rho^{\rA\rB}) \geq f(w) = \min_{\{{M^{*\rV\rB}_{a\pm}}\}_a,\rho^\rB_\pm} \quad \tr[\rho^\rB_-] \nonumber \\
\text{s.t. }&\quad w = \sum_{a,x}\tr[({\omega_x}^{\rV}\otimes F_{a|\omega_x}^\rB)(M_{a+}^{*\rV\rB} - M_{a-}^{*\rV\rB})] \nonumber \\
&\quad \sum_a M_{a\pm}^{*\rV\rB} = \openone^{\rV} \otimes \rho^\rB_\pm \\
&\quad M_{a\pm}^{*\rV\rB} \geq 0 \quad \forall a\nonumber 
\end{align}
This optimization problem is equivalent to \eqref{negtel}. It looks for the state with minimum negativity that could have led to the given violation of the non-classical teleportation witness.

\section{Estimating entanglement robustness from teleportation experiments}\label{rob}


Entanglement negativity is not the only entanglement measure that can be estimated from a teleportation experiment. In this section we turn our attention to entanglement robustness and show how it can be inferred from full data accessible in a teleportation experiment. Moreover, we show that the lower bound can have a meaning of a quantifier of non-classicality of teleportation.

An intuitive way to quantify non-classicality (\eg  entanglement, EPR steering, non-locality, etc.) is in terms of its robustness to noise. Such robustness measures are expressed as the maximal amount of noise which can be added to the given object before it becomes classical. Specifying the type of added noise allows for different types of robustness to be defined. Entanglement robustness \cite{Vidal} is a well known example: for a  bipartite state $\rho^{\rA\rB}$ the entanglement robustness is defined through the following optimization problem
\begin{eqnarray}\label{EntRob}
\epsilon(\rho^{\rA\rB}) &=& \min_{r, \rho_s,\sigma_S} r\\ \nonumber
\textrm{s.t.}&\quad& \frac{\rho^{\rA\rB} + r\rho_s}{1+r}  = \sigma_S\\ \nonumber
&\quad& \sigma_S \in \mathcal{S},
\end{eqnarray}
 
Depending on the properties of $\rho_s$, different types of entanglement robustness can be defined:
\begin{itemize}
\item[$\star$]{\textit{generalized entanglement robustness} \cite{Steiner} $\epsilon_{\textrm{gen}}$, obtained when the only constraint is that $\rho_s$ is a valid quantum state.}
\item[$\star$]{\textit{separable entanglement robustness} $\epsilon_{\textrm{sep}}$, obtained when the state $\rho_s$ is separable.}
\item[$\star$]{\textit{random entanglement robustness} $\epsilon_{\textrm{r}}$, obtained when the state $\rho_s$ is maximally mixed $\rho_s = \frac{\mathds{1}}{d^2}$.}
\end{itemize}
Based on the inclusion relations between the sets of states to which $\rho_s$ belongs, it follows that
\begin{equation*}
\epsilon_{\textrm{gen}} \leq \epsilon_{\textrm{sep}} \leq \epsilon_{\textrm{r}}.
\end{equation*}
Anticipating its role in estimating entanglement robustness let us define teleportation robustness. The central object in a teleportation experiment is the teleportation assemblage $\{\sigma_{a\vert \omega_x}\}_{a,x}$. Robustness of teleportation represents the maximal proportion of a ``noise assemblage" with which $\{\sigma_{a\vert \omega_x}\}_{a,x}$ can be mixed before it becomes classical:
\begin{align} \label{TelRob}
\begin{split}
&\tau (\{\sigma_{a\vert \omega_x}\}) = \max_{r, \{\bar{\sigma}_{a\vert \omega_x}\}, \{M^*_a\}} r \\ 
  \textrm{s.t.} &\quad \frac{\sigma_{a\vert \omega_x}^\rB + r\bar{\sigma}_{a\vert \omega_x}^\rB}{1+r}  = \textrm{tr}_\rV [{M^*_a}^{\rV\rB}(\omega_x^\rV \otimes \mathds{1}^\rB)], \\
&\quad \sum_a {M^*_a}^{\rV\rB} = \mathds{1}^\rV\otimes \frac{\sum_a \sigma_{a\vert \omega_x}^\rB + r\sum_a\bar{\sigma}_{a\vert \omega_x}^\rB}{1+r}, \\ 
&\quad  {M^*_a}^{\rV\rB} \in \mathcal{S}  \qquad \forall a.
\end{split}
\end{align}
The constraints on the `noise assemblage' $\{\bar{\sigma}_{a\vert \omega_x}\}_{a,x}$ determine different types of teleportation robustness:
\begin{itemize}
\item [$\star$] \textit{generalized teleportation robustness} $\tau_{\textrm{gen}}$, obtained when the only constraint on $\{\bar{\sigma}_{a\vert \omega_x}\}_{a,x}$ is that it is allowed by quantum theory.
\item [$\star$] \textit{classical teleportation robustness} $\tau_{\textrm{cl}}$, obtained when $\{\bar{\sigma}_{a\vert \omega_x}\}_{a,x}$ describes classical teleportation.
\item [$\star$] \textit{random teleportation robustness} $\tau_{\textrm{r}}$, obtained when each element of $\{\bar{\sigma}_{a\vert \omega_x}\}_{a,x}$ is proportional to the maximally mixed state.
\end{itemize}

Each type of teleportation robustness places a lower bound on the corresponding type of entanglement robustness of the shared state. This is to be expected, since it has previously been shown that it is possible to place lower bounds on different entanglement quantifiers of the shared state in a measurement-device-independent manner \cite{SCS,Rosset}. Since we have already established teleportation as corresponding to the one-sided measurement-device-independent scenario, it seems natural that teleportation quantifiers are also related to the entanglement quantifiers of the shared state $\rho^{\rA\rB}$ (see (\ref{e:unnstates})). Moreover, we can show that if Alice applies a full Bell state measurement and the set of input states is tomographically complete, then each teleportation robustness equals the corresponding entanglement robustness of the shared state $\rho^{\rA\rB}$. Since the proofs are similar in spirit to those used for lower bounding entanglement negativity we leave the detailed proofs for the appendix.

\section{Teleportation weight and best separable approximation}\label{weight}

In the previous section we saw that the lower bound on entanglement robustness inferred from a teleportation experiment can be seen as a teleportation quantifier. In this section we start from the opposite direction, i.e. we first define a teleportation quantifier and then we show that it puts a lower bound on the corresponding entanglement measure.
\begin{figure}[t!]
\centering
\includegraphics[width=\columnwidth]{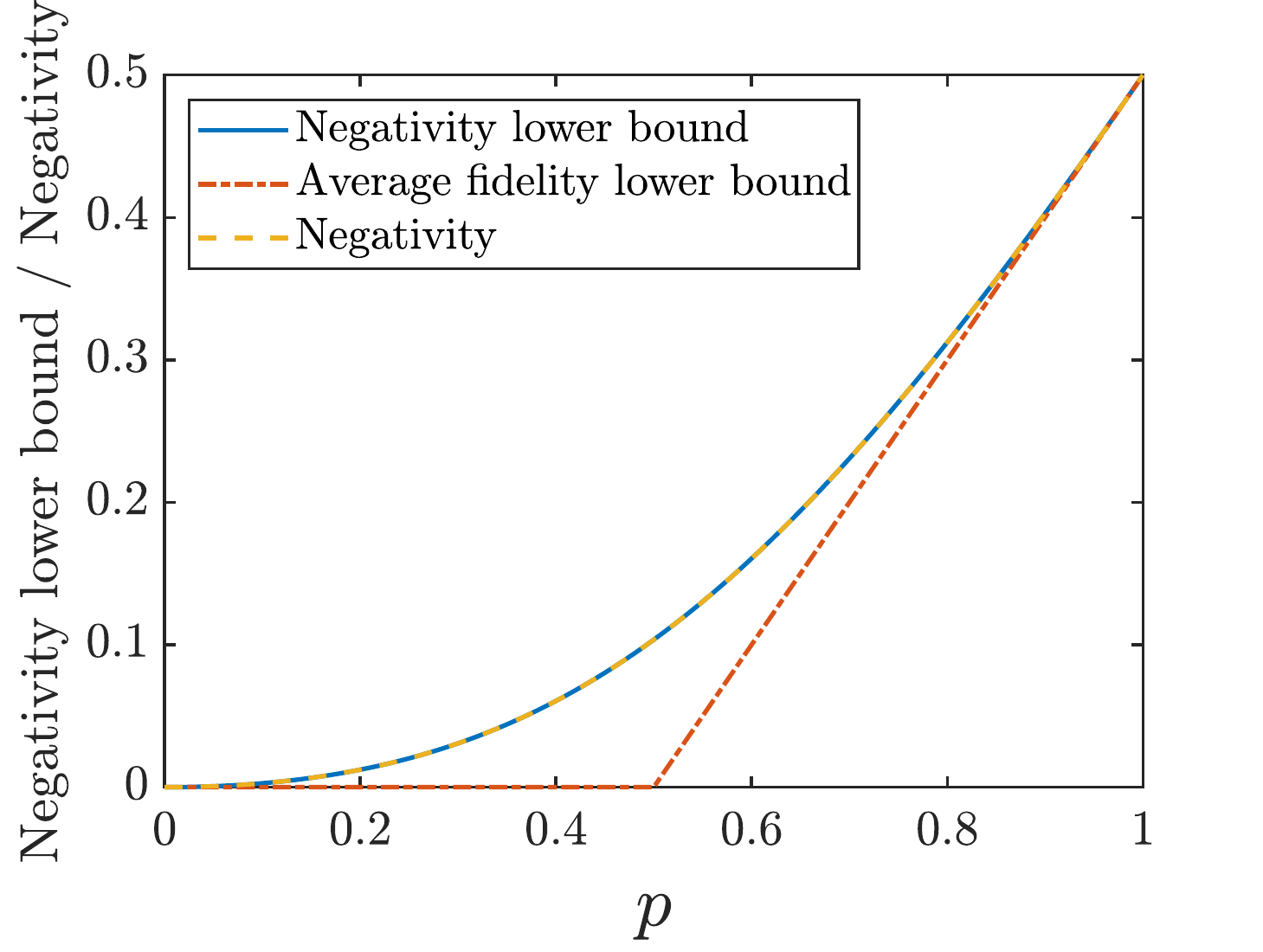}
\caption{Teleportation-based lower bound on the negativity, and negativity of the state $p\ket{\Phi^+}\bra{\Phi^+} + (1-p)\ket{01}\bra{01}$. Alice performs a full Bell State Measurement and uses a tomographically complete set of input states. In this case, the lower bound on the negativity is tight. }
\label{fig:tel neg}
\end{figure}
An operationally meaningful teleportation quantifier different to robustness-based quantifiers is the teleportation weight. Any teleportation assemblage can be written as a convex combination of an assemblage obtained via classical teleportation and a non-classical one. The minimal proportion of the non-classical teleportation assemblage defines the teleportation weight. It can be seen as an analogue to the best Separable approximation \cite{BSA}, EPR2 decomposition \cite{EPR2}, and the steering weight \cite{St}. Mathematically, we define the teleportation weight in the following way
\begin{align} \label{tw}
\begin{split}
\textrm{TW}(&\{\sigma_{a\vert \omega_x}^\rB\}) = \min_{p,\{\bar{M}_{a}\},\{\tilde{M}_{a}\}} \quad  p  \\
\textrm{s.t.} &\quad \sigma_{a\vert \omega_x}^\rB = \textrm{tr}_\rV\left[\left(p\tilde{M}_{a}^{\rV\rB} + \left(1-p\right)\bar{M}_{a}^{\rV\rB}\right)\omega_x^\rV \otimes \mathds{1}^\rB\right] \\
&\quad\sum_a \textrm{tr}_\rV\left[\tilde{M}_{a}^{\rV\rB}(\omega_x^\rV \otimes \mathds{1}^\rB)\right] = \\ &\quad \quad   = \sum_a \textrm{tr}_\rV\left[\tilde{M}_{a}^{\rV\rB}(\omega_{x'}^\rV \otimes \mathds{1}^\rB)\right], \quad \forall x,x', \\
&\quad\bar{M}_{a}^{\rV\rB} \geq 0, \quad \bar{M}_{a}^{\rV\rB} \in \mathcal{S}, \quad \forall a, \\ 
&\quad\left(\tilde{M}_{a}^{*\rV\rB}\right)^{T_\rV} \geq 0, \quad \forall a.
\end{split}
\end{align}
In this definition the channel operators $\bar{M}_{a}^{\rV\rB}$ describe classical teleportation, which is why they have to be positive and separable, while $\tilde{M}_{a}^{\rV\rB}$ are channel operators corresponding to non-classical teleportation, satisfying instead the constraint of the positivity of the partial transpose (see   \eqref{tilrho1}). Non-zero teleportation weight witnesses that teleportation is non-classical, which in turn means that the state Alice and Bob share is entangled. When the set of input states is tomographically complete and the state Alice and Bob share is maximally entangled the teleportation weight must be equal to $1$. Moreover, any pure entangled shared state with tomographically complete set of input states has maximal teleportation weight.\\ 

Just as teleportation robustness quantifiers can be seen to provide bounds on the corresponding entanglement robustness quantifiers, so too does the teleportation weight of the teleportation assemblage $\{\sigma_{a|\omega_x}\}_{a,x}$  place a lower bound on the best separable approximation of the state $\rho^{\rA\rB}$. The best separable approximation of a bipartite state $\rho^{\rA\rB}$ is a monotone which says how much of a separable state is contained in the state $\rho^{\rA\rB}$  and is  defined as 
\begin{align}\label{BSA}
\begin{split}
\epsilon_{\BSA}(\rho^{\rA\rB}) = &\min_{p,\rho_s,\sigma_S}p\\
\textrm{s.t.} \quad & \rho^{\rA\rB} = p\rho_s + (1-p)\sigma_S,\\
&\sigma_S \in \mathcal{S},
\end{split}
\end{align}

\begin{figure}[t!]
\centering
\includegraphics[width=\columnwidth]{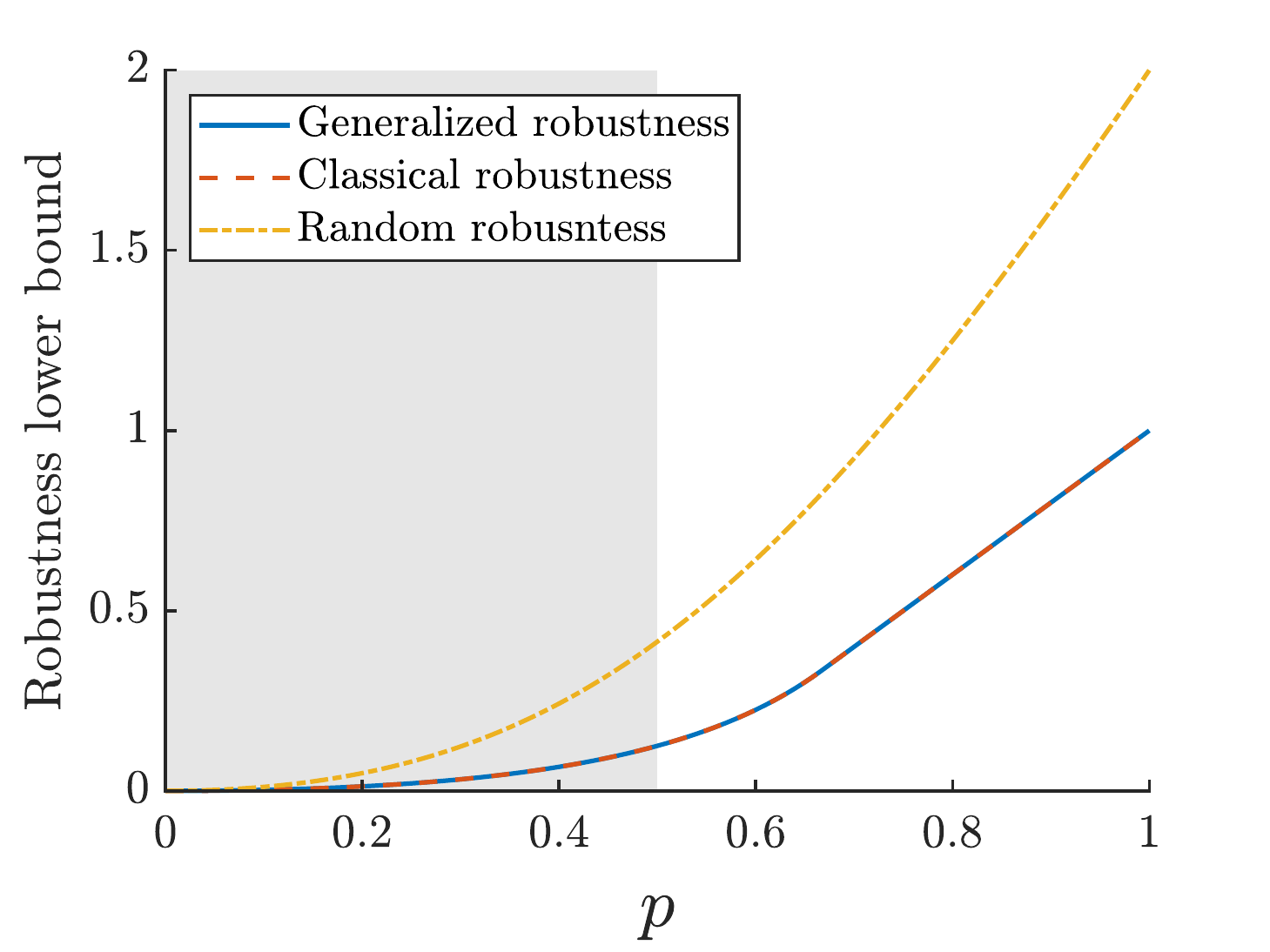}
\caption{Teleportation robustnesses for the state $\rho = p\ket{\Phi^+}\bra{\Phi^+} + (1-p)\ket{01}\bra{01}$. The set of quantum inputs consists of all eigenstates of three Pauli operators. For this particular teleportation assemblage the generalised and classical teleportation robustness coincide. We can see that even when the average teleportation fidelity is smaller than $2/3$ the robustness quantifiers are larger than zero, demonstrating non-classical teleportation.}
\label{fig:Robust}
\end{figure}
We leave the proof that $\textrm{TW}(\{\sigma_{a\vert \omega_x}^\rB\}) \leq \epsilon_{\BSA}(\rho^{\rA\rB})$ for the appendix. We also show in the appendix that in case Alice has a tomographically complete set of input states and performs a Bell state measurement, the resulting teleportation assemblage has teleportation weight equal to the best separable approximation of the state Alice shares with Bob.

\section{Examples}

In this section we present several examples of the methods presented in this work. An online notebook reproducing all of the figures presented in this section can be found at \cite{code}  \\

\textit{Negativity estimation}.-- In Fig.~\ref{fig:tel neg} we plot the teleportation-based lower bound on the negativity for the state $p\ket{\Phi^+}\bra{\Phi^+} + (1-p)\ket{01}\bra{01}$, when Alice performs a full Bell State Measurement and uses a tomographically complete set of input states (eigenstates of the Pauli operators). We see that, as expected, the lower bound is tight, and equal to the negativity of the state. 


\textit{Robustness estimations}.-- In Fig.~\ref{fig:Robust} we show a graph showing different types of teleportation robustness, i.e. lower bounds to the corresponding types of entanglement robustustness of teleportation assemblages produced by using state $\rho = p\ket{\Phi^+}\bra{\Phi^+} + (1-p)\ket{01}\bra{01}$.

\begin{figure}[t!]
\centering
\includegraphics[width=\columnwidth]{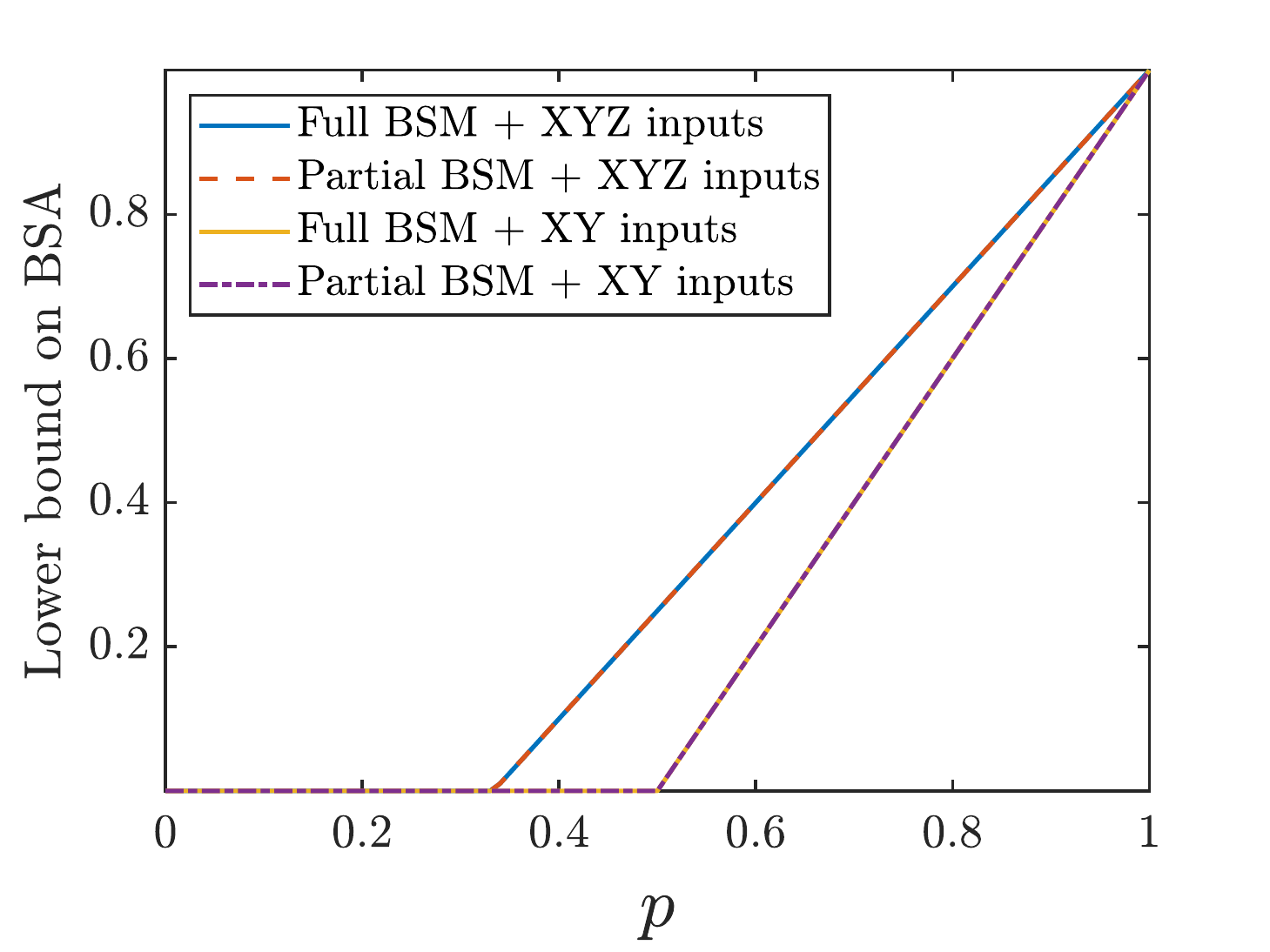}
\caption{Teleportation weight for different scenarios involving the state $p\ket{\Phi^+}\bra{\Phi^+} + (1-p)\frac{\mathds{1}}{4}$: Alice either performs a full or partial Bell State Measurement,  and uses either a tomographically complete set of inputs (eigenstates of $X$,$Y$ and $Z$), or a tomographically incomplete set of measurements (eigenstates of $X$ and $Z$. The teleportation weight is insensitive to the choice of measurements for both sets of inputs, indicating that it is only the conclusive events (corresponding to POVM elements that are entangled) that count.}
\label{fig:TW1}
\end{figure}

\textit{Teleportation weight}.-- The teleportation weight of the state $p\ket{\Phi^+}\bra{\Phi^+} + (1-p)\frac{\mathds{1}}{4}$ corresponding to different scenarios (\ie different sets of input states) is presented in Fig. \ref{fig:TW1}. We see that the teleportation weight for a tomographically complete set of input states is larger than zero whenever $p > \frac{1}{3}$, which is the separability bound for Werner states. This does not change even if Alice does not apply the full Bell state measurements, but projects only onto one of the Bell states (\ie a partial Bell state measurement). When the set of input states consists of eigenstates of two Pauli observables, non-classical teleportation is detected only when $p > \frac{1}{2}$. 

\begin{figure}[t!]
\centering
\includegraphics[width=\columnwidth]{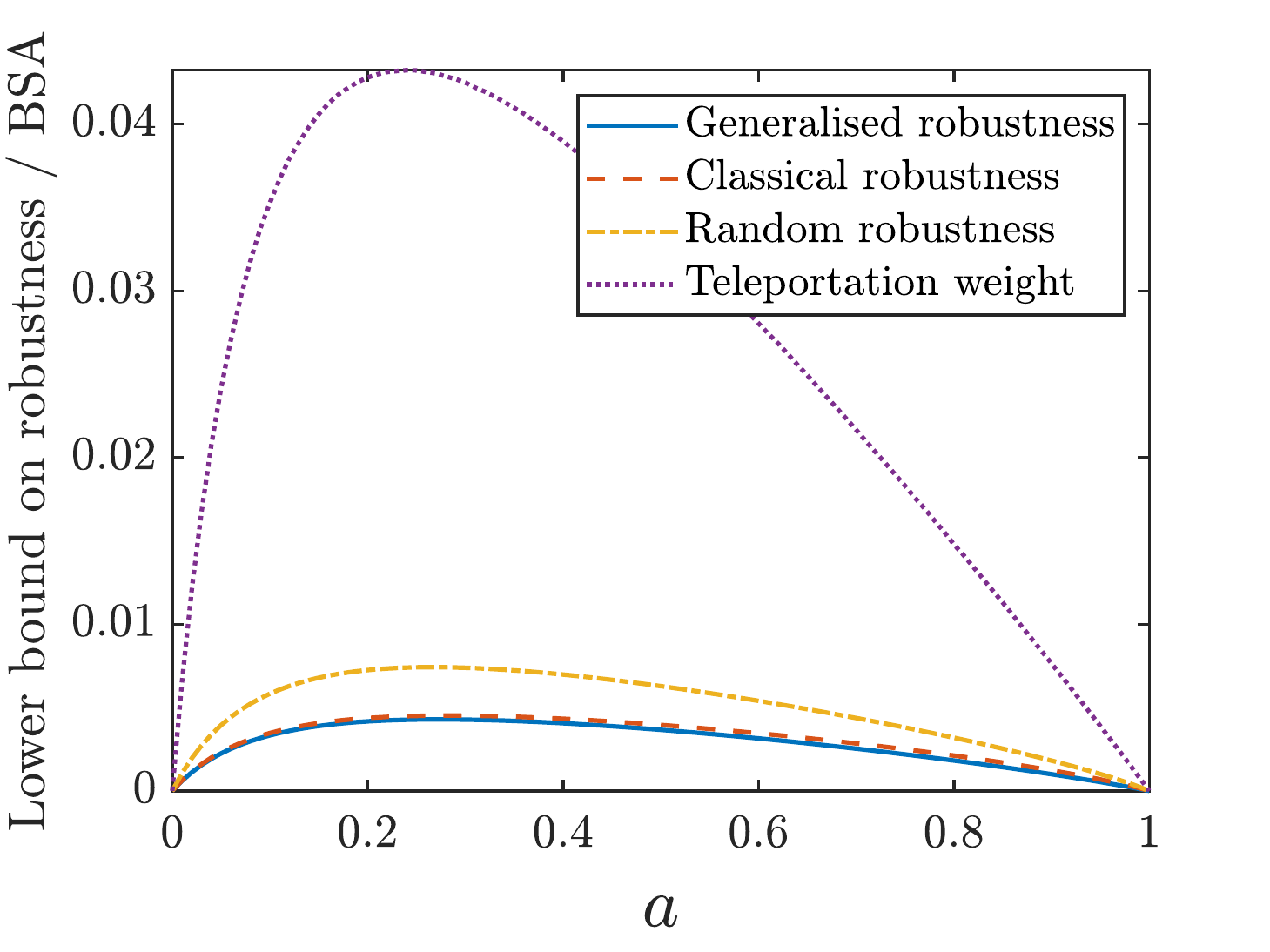}
\caption{Dependence of the teleportation quantifiers introduced here on parameter $a$ a the Horodecki state, using a tomographically complete set of input states (chosen randomly to produce this plot), and a partial Bell State Measurement. For all values of $a \neq 0$ or $1$, non-classical teleportation is demonstrated. Separability of the channel operators was relaxed to the requirement of having a $2$-symmetric PPT extension \cite{Doherty}. }
\label{fig:HorodeckiTW}
\end{figure}

\textit{Bound entangled states}.-- One of the most striking new insights resulting from using all the observable data in a teleportation experiment is that all entangled states can be used to certify non-classical teleportation. Previously, all bound entangled states were considered to be useless for teleportation. One of the most famous  examples of bound entangled states is the Horodecki state \cite{HorodeckiState}:
\be\label{eq:Horodecki}
\rho_H = \frac{1}{8a + 1}
\begin{pmatrix}
a & 0 & 0 & 0 & a & 0 & 0 & 0 & a \\
0 & a & 0 & 0 & 0 & 0 & 0 & 0 & 0 \\
0 & 0 & a & 0 & 0 & 0 & 0 & 0 & 0 \\
0 & 0 & 0 & a & 0 & 0 & 0 & 0 & 0 \\
a & 0 & 0 & 0 & a & 0 & 0 & 0 & a \\
0 & 0 & 0 & 0 & 0 & a & 0 & 0 & 0 \\
0 & 0 & 0 & 0 & 0 & 0 & \frac{1+a}{2} & 0 & \frac{\sqrt{1-a^2}}{2} \\
0 & 0 & 0 & 0 & 0 & 0 & 0 & a & 0 \\
a & 0 & 0 & 0 & a & 0 & \frac{\sqrt{1-a^2}}{2} & 0 & \frac{1+a}{2}
\end{pmatrix},
\ee
for values $a \in (0,1)$. The dependence of the teleportation weight of the teleportation assemblage obtained by using the Horodecki state with parameter $a$ is given on Fig. \ref{fig:HorodeckiTW}. The set of input states is chosen to be tomographically complete and a partial Bell state measurements is performed ($M_1^{\rV\rA} = \ket{\Phi^+}\bra{\Phi^+}$, $M_2^{\rV\rA} = \mathds{1} - M_1^{\rV\rA}$). The teleportation weight of the Horodecki state is small in value, but we observed that other bound entangled states give higher weights, even maintaining a partial Bell state measurement. For example,  the ``pyramid" UPB (unextendable product bases) state \cite{UPB}, with tomographically complete set of inputs, has teleportation weight equal to $0.2350$.

\section{Conclusion}

In this work we have discussed the estimation of entanglement from teleportation experiments and closely related problem of quantifying non-classical teleportation. Taking into account the central role of quantum teleportation in various quantum information protocols, it is very important to improve our understanding of the role of entanglement in the protocol, both qualitatively and quantitatively. We showed that by using the full data available in an experiment we can put lower bounds on several entanglement measures, such as entanglement negativity, robustness of entanglement and the best separable approximation. Importantly, in accordance with our previous work \cite{CSS} we have shown that we can estimate non-zero entanglement even from teleportation experiments in which the average fidelity of teleportation falls below the classical limit. The inferred lower bound appears to be tight in case when Alice applies a full Bell state measurements and the set of input states is tomographically complete.\\

It is worth mentioning that lower bounds on different entanglement measures can actually be seen as teleportation quantifiers. We thus paid additional attention to operationally meaningful  robustness based teleportation quantifiers, and the teleportation weight. \\

There are several possible directions for future research. One would be to identify other known entanglement measures which can be estimated from a teleportation experiment. A more demanding task would be to mirror the protocol of certifying and estimating non-classicality of teleportation to some other protocols which use teleportation as a sub-routine. Finally, while all our results regard finite-dimensional teleportation it would be interesting to understand what can be said about continuous variable teleportation.\\

\section{acknowledgements}
This work was supported by the Ram\'on y Cajal fellowship, Spanish MINECO (QIBEQI FIS2016-80773-P and Severo Ochoa SEV-2015-0522), the AXA Chair in Quantum Information Science, Generalitat de Catalunya (SGR875 and CERCA Programme), Fundaci\'{o} Privada Cellex and ERC CoG QITBOX, and a Royal Society University Research Fellowship (UHQT).

\begin{appendix}

\section{Lower bounds on entanglement robustness}

The entanglement present in the shared state $\rho^{\rA\rB}$ is a necessary resource for non-classicality of teleportation assemblage $\{\sigma_{a|\omega_x}\}_{a,x}$. Given a robustness-based non-classicality measures for both the state $\rho^{\rA\rB}$ and the teleportation assemblage $\{\sigma_{a|\omega_x}\}_{a,x}$, it is instructive to compare their values. 

In this paper we prove the following relations 
\begin{align}
&\tau_{\textrm{\g}}(\{\sigma_{a|\omega_x}\}) \leq \epsilon_{\textrm{gen}}(\rho^{\rA\rB}), \label{liana}\\
&\tau_{\textrm{cl}}(\{\sigma_{a|\omega_x}\}) \leq \epsilon_{\textrm{sep}}(\rho^{\rA\rB}); \label{raeghar}\\
&\tau_{\textrm{r}}(\{\sigma_{a|\omega_x}\}) \leq \epsilon_{\textrm{r}}(\rho^{\rA\rB}). \label{elia}
\end{align}
All inequalities are saturated when Alice teleports a tomographically complete set of inputs and performs a Bell state measurement.

\subsubsection{Generalized teleportation robustness}

Let us examine in detail the generalized teleportation robustness of a teleportation assemblage $\{\sigma_{a|\omega_x}\}_{a,x}$, obtained by using the measurement $\{M_a\}$ on a shared state $\rho^{\rA\rB}$, and a set of input states $\{\omega_x\}_x$. This quantity is given by \eqref{TelRob} when the only constraint on $\{\bar{\sigma}_{a|\omega_x}\}_{a,x}$ is that it is admissible by quantum theory: 
\begin{align} \label{g}
\begin{split}
&\tau _{\textrm{gen}}(\{\sigma_{a\vert \omega_x}\}) = \min_{r,\{M^*_a\},\{\bar{\sigma}_{a|\omega_x}\}} r\\ 
\textrm{s.t.} &\quad   \frac{\sigma_{a\vert \omega_x}^{\rB} + r\bar{\sigma}_{a|\omega_x}^{\rB}}{1+r} = \textrm{tr}_{\rV}\left[{M_a^*}^{\rV\rB}(\omega_x^V \otimes \mathds{1}^{\rB})\right],\\ 
&\quad  \bar{\sigma}_{a|\omega_x}^{\rB} \in T_{q},\\ 
&\quad  \sum_a {M_a^*}^{\rV\rB} = \mathds{1}^{\rV}\otimes\frac{\sum_a\sigma_{a|\omega_x} + r\sum_a\bar{\sigma}_{a|\omega_x}}{r+1}, \\ 
&\quad  {M^*_a}^{\rV\rB} \geq 0, \quad  {M^*_a}^{\rV\rB} \in \mathcal{S}, \quad \forall a;
\end{split}
\end{align}
where $T_q$ is the set of teleportation assemblages admissible by quantum theory -- \ie those which can arise by performing a quantum measurement on a shared quantum state (this requirement will be made more explicit in the appendix). The set of constraints given above imposes that the mixture of the observed assemblage $\{\sigma_{a\vert \omega_x}\}_{a,x}$ with some other hypothetical assemblage $\{\bar{\sigma}_{a\vert \omega_x}\}_{a,x}$ can be simulated classically. We note that all constraints can be written in a linear form -- with a suitable relaxation of the set of separable operators this problem is readily solved by using semidefinite programming optimization (SDP) \cite{SDP}. The only non-trivial constraint regards the characterization of the set $T_q$, but in the next section we will show that membership in such a set also can be imposed as a semidefinite programming constraint.\\

From the definition of the generalized entanglement robustness we know that 
\begin{equation}\label{ned}
\frac{\rho^{\rA\rB} + \epsilon_{\g}(\rho^{\rA\rB})}{1+\epsilon_{\g}(\rho^{\rA\rB})}  = \sigma_S
\end{equation}
for some state $\rho_s$ and a separable state $\sigma_S$. By tensoring with $\omega_x^{\rV}$ and  applying measurement $\{M_a^{\rV\rA}\}_a$, \eqref{ned} becomes
\begin{align*}
&\quad \frac{1}{1+\epsilon_{\g}(\rho^{\rA\rB})}\tr_{\rV\rA}\left[\left(M_a^{\rV\rA}\otimes \mathds{1}^{\rB}\right)\left(\omega_x^{\rV}\otimes\rho^{\rA\rB}\right)\right]  \\ &+ \frac{\epsilon_{\g}(\rho^{\rA\rB})}{1+\epsilon_{\g}(\rho^{\rA\rB})}\tr_{\rV\rA}\left[\left(M_a^{\rV\rA}\otimes \mathds{1}^{\rB}\right)\left(\omega_x^{\rV}\otimes\rho_{s}^{\rA\rB}\right)\right]\\& = \tr_{\rV\rA}\left[\left(M_a^{\rV\rA}\otimes \mathds{1}^{\rB}\right)\left(\omega_x^{\rV}\otimes\sigma_S^{\rA\rB}\right)\right]
\end{align*}
for every $x$ and $a$. This is equivalent to 
\begin{eqnarray}\label{jon}
&\dfrac{\sigma_{a|\omega_x}+\epsilon_{\g}(\rho^{\rA\rB})\bar{\sigma}_{a|\omega_x}}{1+\epsilon_{\g}(\rho^{\rA\rB})} = \tr_{\rV}\left[{M_a^*}^{\rV\rB}\left(\omega_x^{\rV}\otimes \mathds{1}^{\rB}\right)\right]
\end{eqnarray}
where 
\begin{equation*}
\bar{\sigma}_{a|\omega_x} = \tr_{\rV\rA}\left[\left(M_a^{\rV\rA}\otimes \mathds{1}^{\rB}\right)\left(\omega_x^{\rV}\otimes\rho_{s}^{\rA\rB}\right)\right],
\end{equation*}
is an arbitrary teleportation assemblage (for an arbitrary state $\rho_s$). At the same time, 
\begin{align}\label{hodor}
{M_a^*}^{\rV\rB} = \tr_{\rA}\left[\left(M_a^{\rV\rA}\otimes \mathds{1}^{\rB}\right)\left(\mathds{1}^{\rV}\otimes \sigma_S^{\rA\rB}\right)\right]
\end{align}
is separable when $\sigma_S$ is separable. Furthermore
\begin{align}\label{cat}
\begin{split}
\sum_a {M_a^*}^{\rV\rB} &= \tr_{\rA}\left[\left(\sum_aM_a^{\rV\rA}\otimes \mathds{1}^{\rB}\right)\left(\mathds{1}^{\rV}\otimes \sigma_S^{\rA\rB}\right)\right]\\ 
&= \tr_{\rA}\left(\mathds{1}^V\otimes \sigma_S^{\rA\rB}\right)\\ 
&= \mathds{1}^{\rV}\otimes \frac{\tr_{\rA}{\rho^{\rA\rB}} + \epsilon_{\g}(\rho^{\rA\rB})\tr_{\rA}{\rho_s^{\rA\rB}}}{1+\epsilon_{\g}(\rho^{\rA\rB})}\\ 
&= \mathds{1}^{\rV}\otimes \frac{\sum_a\sigma_{a|\omega_x} + \epsilon_{\g}(\rho^{\rA\rB})\sum_a{\bar{\sigma}_{a|\omega_x}}}{1+\epsilon_{\g}(\rho^{\rA\rB})}
\end{split}
\end{align}
where the second line follows from the completeness relation $\sum_aM_a = \mathds{1}$, the third line follows from \eqref{ned}, and the last line is obtained by using the definitions of $\sigma_{a|\omega_x}$ and $\bar{\sigma}_{a|\omega_x}$.

 From \eqref{jon}, \eqref{cat} and the separability of ${M_a^*}^{\rV\rB}$, it follows that  $\epsilon_{\g}(\rho^{\rA\rB})$ satisfies all the constraints from \eqref{g}. Since the mixing teleportation assemblage $\bar{\sigma}_{a|\omega_x}$ did not have any special properties, apart from being realizable in quantum theory, the generalized teleportation robustness  of $\{\sigma_{a|\omega_x}\}_{a,x}$ cannot be bigger than $\epsilon_{\g}(\rho^{\rA\rB})$:
\begin{equation}
\tau_{\textrm{\g}}(\{\sigma_{a|\omega_x}\}) \leq \epsilon_{\textrm{gen}}(\rho^{\rA\rB}).
\end{equation}

\subsubsection{Classical teleportation robustness}

The classical teleportation robustness $\tau_{\textrm{cl}}(\cdot)$ is defined by (\ref{TelRob}) with the additional constraint that the mixing teleportation assemblage describes classical teleportation. Such a teleportation assemblage is characterised by positive and separable channel operators $\bar{M}_a^{\rV\rB}$ as shown in (\ref{LocModChOP}) and (\ref{savnik}). With these constraints, the classical teleportation robustness can be expressed as the following optimization problem
\begin{eqnarray} \nonumber
&\tau_{\textrm{cl}}&(\{\sigma_{a\vert \omega_x}\}) = \min_{\{r_a\},\{\bar{M}_a\},\{{M^*_a}\}} r\\ \nonumber
\textrm{s.t.} &\quad &  \frac{\sigma_{a\vert \omega_x}^{\rB} + r\bar{\sigma}^{\rB}_{a|\omega_x}}{1+r} = \textrm{tr}_{\rV}[M_a^{*\rV\rB}(\omega_x^{\rV} \otimes \mathds{1}^{\rB})];\\ \nonumber
&\quad & \bar{\sigma}_{a|\omega_x}^{\rB} = \tr_{\rV}\left[\left({\bar{M}_a}^{\rV\rB}\right)\left(\omega_x^{\rV}\otimes {\mathds{1}}^{\rB}\right)\right] \quad \forall a,x \\ \label{cl}
&\quad & \bar{M}_a^{\rV\rB}, M_a^{*\rV\rB} \geq 0, \quad \forall a; \\ \nonumber
&\quad & \bar{M}_a^{\rV\rB}, M_a^{*\rV\rB} \in \mathcal{S}, \quad \forall a; \\ \nonumber
&\quad &  \sum_a \bar{M}_a^{\rV\rB} = \mathds{1}^{\rV}\otimes \sum_a{\bar{\sigma}_{a|\omega_x}}; \\ \nonumber
&\quad &  \sum_aM_a^{*\rV\rB} = \mathds{1}^{\rV}\otimes \frac{\sum_a{{\sigma}_{a|\omega_x}} + r\sum_a{\bar{\sigma}_{a|\omega_x}}}{1+r}.
\end{eqnarray}
The classical teleportation robustness mirrors the classical entanglement robustness, $\epsilon_{\textrm{sep}}(\rho^{\rA\rB})$ which satisfies the following equation
\begin{equation}\label{sansa}
\frac{\rho^{\rA\rB}+\epsilon_{\textrm{sep}}(\rho^{\rA\rB})\rho_s}{1+\epsilon_{\cc}(\rho^{\rA\rB})}  = \sigma_S
\end{equation}
where now both states $\rho_s$ and $\sigma_S$ are separable. By tensoring with $\omega_x^{\rV}$ and applying the measurement $\{M_a^{\rV\rA}\}_a$ this becomes 
\begin{equation}\label{arya}
\dfrac{\sigma_{a|\omega_x} + \epsilon_{\textrm{sep}}(\rho^{\rA\rB})\bar{\sigma}_{a|\omega_x}}{1+\epsilon_{\textrm{sep}}(\rho^{\rA\rB})}  = \tr_{\rV}\left[{M_a^*}^{\rV\rB}\left(\omega_x^{\rV}\otimes \mathds{1}^{\rB}\right)\right]
\end{equation}
where now the mixing assemblage can be expressed in terms of separable channel operators ,
${\bar{M}_a}^{\rV\rB}$
\begin{align*}
\bar{\sigma}_{a|\omega_x} &= \tr_{\rV}\left[{\bar{M}_a}^{\rV\rB}\left(\omega_x^{\rV}\otimes \mathds{1}^{\rB}\right)\right];\\
{\bar{M}_a}^{\rV\rB} &= \tr_{\rA}\left[\left(M_a^{\rV\rA}\otimes \mathds{1}^{\rB}\right)\left(\mathds{1}^{\rV}\otimes \rho_s^{\rA\rB}\right)\right];\\
\sum_a{\bar{M}_a}^{\rV\rB} &= \mathds{1}^{\rV}\otimes \sum_a\bar{\sigma}_{a|\omega_x}
\end{align*}
The channel operators ${M_a^*}^{\rV\rB}$ remain separable and still satisfy relation \eqref{cat}. Together with \eqref{arya} and the separability of ${\bar{M}_a}^{\rV\rB}$, this implies that $\epsilon_{\textrm{sep}}(\rho^{\rA\rB})$ satisfies all of the constraints of the classical teleportation robustness (of the teleportation assemblage $\{\sigma_{a|\omega_x}\}_{a,x}$), leading to:
\begin{equation}
\tau_{\textrm{cl}}(\{\sigma_{a|\omega_x}\}) \leq \epsilon_{\textrm{sep}}(\rho^{\rA\rB}).
\end{equation}
\subsubsection{Random teleportation robustness}

In \cite{CSS} the random teleportation robustness was introduced  as a special case of (\ref{TelRob}) with the constraint $\bar{\sigma}_{a| \omega_x} = \frac{1}{|o|}\frac{\mathds{1}}{d}$, where $|o|$ is the number of outcomes of Alice's measurement. Here we consider a more general definition,  defined as the solution to the following optimization problem
\begin{align}\label{r}
\begin{split}
\tau_{\textrm{r}}(\{\sigma_{a|\omega_x}\})  &= \min_{r, \{p(a)\},\{M^*_a\}} r\\ 
\textrm{s.t.} &\quad \frac{\sigma_{a\vert \omega_x}^{\rB} + r p(a)\frac{\mathds{1}^{\rB}}{d}}{1+r} = \textrm{tr}_V [{M^*_a}^{VB}(\omega_x^V \otimes \mathds{1}^B)], \\ 
&\quad \sum_ap(a) = 1, \quad {M^*_a}^{\rV\rB} \geq 0, \quad {M^*_a}^{\rV\rB} \in \mathcal{S} \\ 
&\quad  \sum_a{M^*_a}^{\rV\rB} = \mathds{1}^V \otimes \frac{\rho^{\rB} + r\frac{\mathds{1}^B}{d}}{1+r} .
\end{split}
\end{align}
Recall that the random entanglement robustness $\epsilon_{\textrm{r}}(\rho^{\rA\rB})$ satisfies
\begin{equation}\label{bran}
\frac{\rho^{\rA\rB} + \epsilon_{\rr}(\rho^{\rA\rB}) \mathds{1}/d^2}{1+\epsilon_{\rr}(\rho^{\rA\rB})} = \sigma_S.
\end{equation}
Analogously to the previous cases this equation implies
\begin{equation*}
\frac{\sigma_{a\vert \omega_x}^{\rB} + \epsilon_{\rr}(\rho^{\rA\rB})\mathds{1}/d}{1+\epsilon_{\rr}(\rho^{\rA\rB})} = \textrm{tr}_V [{M^*_a}^{\rV\rB}(\omega_x^V \otimes \mathds{1}^B)],
\end{equation*}
where 
\begin{align}\label{benjen}
p(a) = \tr\left[M_a^{\rV\rA}\left(\omega_x^{\rV}\otimes \frac{\mathds{1}^A}{d}\right)\right]
\end{align}
and ${M^*_a}^{\rV\rB}$ satisfies \eqref{hodor} and \eqref{cat} when $\sum_a\bar{\sigma}_{a|\omega_x} = \frac{\mathds{1}}{d}$. Note that \eqref{benjen} confirms that $\sum_ap(a) = 1$. We have confirmed that $\epsilon_{\rr}(\rho^{\rA\rB})$ satisfies all the constraints from \eqref{r}, thus represents an upper bound to the random teleportation robustness
\begin{equation}
\tau_{\textrm{r}}(\{\sigma_{a|\omega_x}\}) \leq \epsilon_{\textrm{r}}(\rho^{\rA\rB}).
\end{equation}

\subsection{Tightness of the lower bound}

In \cite{CSS}, using a form of random teleportation robustness, it was proven that every entangled state leads to non-classical teleportation. This was done by proving that the random entanglement robustness of $\rho^{\rA\rB}$ is proportional to the random teleportation robustness of the teleportation assemblage $\Tel$ that results from applying a (full or partial) Bell state measurement on $\rho^{\rA\rB}$ and a tomographically complete set of input states $\{\omega_x\}_x$. 

In this appendix we prove that such an equivalence is true for each kind of teleportation/entanglement robustness and, moreover, whenever Alice applies the full Bell state measurement the inequalities \eqref{liana} -- \eqref{elia} are saturated.

Before doing so, let us state a lemma which will be repeatedly used in these proofs.

\begin{lem}\label{BSMTR}
Every element of a teleportation assemblage $\{\sigma_{a|\omega_x}\}_{a,x}$ resulting from an arbitrary measurement $M_a^{VA}$ and a shared state $\rho^{\rA\rB}$ could have also been obtained, up to a multiplicative factor, by post-selecting on the measurement outcome ${\Phi^+}^{\rV\rA} = \ket{\Phi^+}\bra{\Phi^+}^{\rV\rA}$ applied to a suitable state ${\rho'_a}^{\rA\rB}$.
\end{lem}

\textit{Proof.}
The identity \eqref{MaxEntIdn} allows any member of a teleportation assemblage $\{\sigma_{a|\omega_x}\}_{a,x}$ to be written in the following way
\begin{eqnarray} \nonumber
&\quad& \sigma_{a\vert \omega_x}^{\rB} = \textrm{tr}_{\rV\rA}\left[ (M_a^{\rV\rA}\otimes \mathds{1}^{\rB})(\omega_x^{\rV} \otimes \rho^{\rA\rB}) \right]\\ \nonumber
&=& d\textrm{tr}_{\rV\rV_1\rA}\left[\left( (\omega_x^{\rV})^T\otimes M_a^{\rV_1\rA} \otimes \mathds{1}^{\rB} \right)\left( {\Phi^+}^{\rV\rV_1}\otimes \rho^{\rA\rB}\right)\right] \\  \nonumber
&=& d\textrm{tr}_{\rV\rV_1\rA}\big[\left (\omega_x^{\rV})^T \otimes \mathds{1}^{\rV_1\rA\rB} \right) \left(  \mathds{1}^{\rV}\otimes M_a^{\rV_1\rA} \otimes \mathds{1}^{\rB} \right)  \times \\   \label{berane} &\quad& \quad \quad \quad \times\left( {\Phi^+}^{\rV\rV_1}\otimes \rho^{\rA\rB}\right)\big]
\end{eqnarray}
In case $M_a^{\rV_1\rA} = {\Phi^+}^{\rV_1\rA}$ the previous equation reduces to:
\begin{eqnarray*}
\sigma_{a\vert \omega_x}^{\rB} = \frac{1}{d}\textrm{tr}_{\rV}\left[\left((\omega_x^{\rV})^T \otimes \mathds{1}^{B} \right)\rho^{\rV\rB}\right]
\end{eqnarray*}
On the other hand, if $M_a^{\rV_1\rA}$ is not a Bell state measurement, \eqref{berane} reduces to
\begin{eqnarray} \nonumber
\sigma_{a\vert \omega_x}^{\rB} &=& d\textrm{tr}_{\rV\rV_1}\left[\left (\omega_x^{\rV})^T \otimes \mathds{1}^{\rV_1\rA\rB} \right){{\rho'_a}}^{\rV\rV_1\rA\rB}\right]\\ \label{4} 
&=& dp(a)\textrm{tr}_{\rV}\left[\left((\omega_x^{\rV})^T \otimes \mathds{1}^{\rB} \right){\rho'_a}^{\rV\rB}\right].
\end{eqnarray}
where 
\begin{equation}\label{pa}
p(a) = \tr[(\mathds{1}^{\rV}\otimes M_a^{\rV_1\rA} \otimes \mathds{1}^{\rB} )( {\Phi^+}^{\rV\rV_1}\otimes \rho^{\rA\rB})],
\end{equation}
and
\begin{equation*}
{\rho_a'}^{\rV\rB} = 
\frac{1}{p(a)}\textrm{tr}_{\rV_1\rA}\left(  \mathds{1}^{\rV}\otimes M_a^{\rV_1\rA} \otimes \mathds{1}^{\rB} \right)\left( {\Phi^+}^{\rV\rV_1}\otimes \rho^{\rA\rB}\right).
\end{equation*}
The state ${\rho_a'}^{\rV\rB}$ can be obtained from the state $\rho^{\rA\rB}$ through a stochastic local operation, which can be seen as a local version of entanglement swapping. To obtain $\rho_a'$ from $\rho$, Alice uses two auxiliary systems in the maximally entangled state, and applies the measurement $M_a$ on one auxiliary system and her share of the state $\rho^{\rA\rB}$. After the measurement she discards the measured systems. 
 
Finally, the expression given in \eqref{4} can be written as:
\begin{eqnarray} \nonumber
\sigma_{a\vert \omega_x}^{\rB} &=& d^2p(a)\textrm{tr}_{\rV\rA}\Big[\left(\omega_x^{\rV} \otimes \mathds{1}^{\rA\rB} \right)\left( {\Phi^+}^{\rV\rA}\otimes {\mathds{1}}^{\rB} \right) \times \\ \nonumber &\quad& \qquad \qquad \times \left( {\mathds{1}}^{\rV}\otimes {\rho_a'}^{\rA\rB}\right)\Big]\\ \label{5} &=& d^2p(a)\textrm{tr}_{\rV\rA}\left[({\Phi^+}^{\rV
\rA}\otimes \mathds{1}^{\rB})(\omega_x^{\rV} \otimes{\rho_a'}^{\rA\rB}) \right],
\end{eqnarray}
which proves the lemma. The multiplicative factor, mentioned in the lemma,  is equal to $d^2p(a)$. $\square$\\

The assumption that Alice applies the Bell state measurement implicitly assumes that the dimension of Alice's reduced state $d_{\rA} = \tr_{\rB}(\rho^{\rA\rB})$ is equal to the dimension of the input state $d$. In the general case, $d_{\rA}$ can be different from $d$ and in \eqref{berane} -- \eqref{5} there is no assumption about $d_{\rA}$.

\subsubsection{Generalized teleportation robustness}

Let us denote by $\tau_{\textrm{gen}}^*(\cdot)$ the generalized teleportation robustness of a teleportation assemblage obtained when Alice performs a Bell state measurement $\{M_a^{\rV\rA}\}$ and the set of input states $\{\omega_x\}_x$ is tomographically complete. In this case the first constraint from \eqref{g} can be rewritten as
\begin{eqnarray}\label{k}
&& \frac{\sigma_{a\vert \omega_x}^{\rB}+r{\bar{\sigma}_{a|\omega_x}}^{\rB}}{1+r} \\ \nonumber &=&
\frac{1}{r+1}\textrm{tr}_{\rV\rA}\left[\left(U_a^{\rA}{\Phi^+}^{\rV\rA}{U_a^{\dagger}}^{\rA}\otimes \mathds{1}^{\rB}\right)\left(\omega_x^{\rV} \otimes \rho^{\rA\rB}\right)\right] \\ \nonumber &\quad& +\frac{1}{r+1}\textrm{tr}_{\rV\rA}\left[\left({\Phi^+}^{\rV\rA}\otimes \mathds{1}^{\rB}\right)\left(\omega_x^{\rV}\otimes d^2p(a){\rho'_a}^{\rA\rB}\right)\right]\\ \nonumber
&=& \frac{1}{d}\textrm{tr}_{\rV}\left[\left(\frac{U_a^{\rV}\rho^{\rV\rB}{U_a^{\dagger}}^{\rV} + rd^2p(a){\rho'_a}^{\rV\rB}}{r+1} \right)^{T_{\rV}}\left(\omega_x^{\rV}\otimes \mathds{1}^{\rB}\right)\right]\\ \nonumber
&=& \textrm{tr}_{\rV}[{M^*_a}^{\rV\rB}(\omega_x^{\rV} \otimes \mathds{1}^{\rB})].
\end{eqnarray}
The second line follows from the assumption that Alice applies a Bell state measurement and $U_a$ are the local unitary transformations that shift between the different Bell states $M_{a} = U_a\Phi^+U_a^{\dagger}$. The third line uses Lemma \ref{BSMTR} and \eqref{5} to re-express the general teleportation assemblage $\{\bar{\sigma}_{a|\omega_x}\}_{a,x}$. To fourth line follows from the identity \eqref{MaxEntIdn}. 

Given that the set of quantum inputs $\{\omega_x\}$ is tomographically complete the last equality implies
\begin{equation}\label{pluzine}
{M^*_a}^{\rV\rB} = \frac{1}{d}\left(\frac{1}{r+1}U_a^{\rV}\rho^{\rV\rB}{U_a^{\dagger}}^{\rV} + \frac{r}{1+r}d^2p(a){\rho'}_a^{\rV\rB}\right)^{T_{\rV}}
\end{equation}
  
 With this in  mind we can once more rewrite the optimization problem (\ref{g}):
\begin{subequations}\label{g1}
\begin{align} \nonumber
&\tau^*_{\textrm{gen}}(\{\sigma_{a\vert \omega_x}\}) = \min_{\{r_a, M^*_a, p(a),\rho'_a\}_a} r \\ \label{g2}
\textrm{s.t.} &\quad   \frac{U_a^{\rV}\rho^{\rV\rB}{U_a^{\dagger}}^{\rV} + r d^2p(a){\rho'}_a^{\rV\rB}}{r+1}  = d\left({M_a^*}^{\rV\rB}\right)^{T_{\rV}},\\  \label{g4}
&\quad  {M^*_a}^{\rV\rB} \geq 0, \quad  {M^*_a}^{\rV\rB} \in \mathcal{S} \quad \forall a, \\  \label{g6}
&\quad  \sum_a p(a){\rho'}_a^{\rV\rB} = \frac{\mathds{1}^{\rV}}{d}\otimes\rho'^{\rB}, \quad \sum_ap(a) = 1,  \\ \label{g3}
&\quad  \sum_a{M^*_a}^{\rV\rB} = \mathds{1}^{\rV}\otimes \frac{\rho^{\rB} + r\rho'^{\rB}}{1+r}, 
\end{align}
\end{subequations}
which resembles the optimization problem defining the generalized entanglement robustness of the state $\rho^{\rV\rB} = \rho^{\rA\rB}$. Indeed the optimization problem (\ref{g1}), for one specific value of $a$, say $a = 0$, for which $U_0 = \mathds{1}$, is similar to (\ref{EntRob}), with the differences being that $d^2p(a)\rho'_a$ and $d{M_a^*}^{VB}$ are not necessarily normalized, and the two additional constraints \eqref{g6} and \eqref{g3}. The constraint (\ref{g2}), for $a=0$, has a solution if  $d\tr{{M^*_0}^{\rV\rB}} = (1+rd^2p(0))/(1+r)$. Taking this into account the constraint can be rearranged in the following way
\begin{eqnarray}\label{banjani}
\frac{\rho^{\rV\rB} + rd^2p(0){\rho'_{0}}^{\rV\rB}}{1+rd^2p(0)} = \frac{1}{\tr{{M^*_0}^{\rV\rB}}}\left({M^*_0}^{\rV\rB}\right)^{T_{\rV}},
\end{eqnarray}
which is now equivalent to the first constraint from (\ref{EntRob}). Thus, the minimal $r$ satisfying this constraint for  separable ${{M^*_0}^{\rV\rB}}$ is equal to $\epsilon_{\textrm{gen}}(\rho^{\rA\rB})/(d^2p(0))$. 

In a similar manner the minimal $r$ satisfying  (\ref{g2}) for $a \neq 0$ and separable ${{M^*_a}^{\rV\rB}}$  is equal to $\epsilon_{\textrm{gen}}(\rho^{\rA\rB})/(d^2p(a))$, since the generalized entanglement robustness is the same for all states which are related by local unitary transformations. Let us, for a moment, suppose that there is at least one $a$ such that $d^2p(a) \geq 1$. Since there are $d^2$ different values of $a$,   (\ref{g6}) implies that for some other value of $a$, say $a= a'$ it must be that $d^2p(a') \leq 1$. But in this case the smallest $r$ satisfying the constraints (\ref{g2}) and (\ref{g4}) for all $a$ must be strictly bigger than $\epsilon_{\textrm{gen}}(\rho^{AB})$. On the other hand, if $d^2p(a) = 1$ for all values of $a$ the 
smallest $r$ satisfying (\ref{g1}) is exactly equal to $\epsilon_{\textrm{gen}}(\rho^{\rA\rB})$. Since $p(a)$ are optimization variables, the minimal $\tau_{\textrm{gen}}$ will be achieved when all $p(a)$s are mutually equal.\\

Finally, we have to make sure that the constraints (\ref{g6}) and (\ref{g3}) are satisfied by the solution $r = \epsilon_{\textrm{gen}}(\rho^{\rV\rB})$. If for for $a=0$ (\ref{g2}) is satisfied for some ${\rho_0'}^{\rV\rB}$ and ${M_0^*}^{\rV\rB}$, for $a \neq 0$ it will be satisfied with the same $r$, ${\rho_a'}^{\rV\rB} = U_a^{\rV}{\rho_0'}^{\rV\rB}{U_a^{\dagger}}^{\rV}$ and ${M_a^*}^{\rV\rB} = U_a^{\rV}{M_0^*}^{\rV\rB}{U_a^{\dagger}}^{\rV}$  implying
\begin{equation*}
\sum_ap(a){\rho_a'}^{\rV\rB} = \frac{\mathds{1^{\rV}}}{d}\otimes {\rho'}^{\rB} 
\end{equation*}
because $\sum_aU_a^{\rV}{\rho_a'}^{\rV\rB}{U_a^{\dagger}}^{\rV} = d\mathds{1}^{\rV}\otimes {\rho'}^{\rB}$. The validity of (\ref{g3}) is verified by summing (\ref{g2}) over all different values of $a$:
\begin{eqnarray*}
\sum_a{M_a^*}^{\rV\rB} &=& \frac{1}{d}\sum_aU_a^{\rV}\frac{\left(\rho^{\rV\rB}\right)^{T_{\rV}} + r\left({\rho'_0}^{\rV\rB}\right)^{T_{\rV}}}{1+r}{U_a^{\dagger}}^{\rV}\\
&=& \mathds{1}^{\rV}\otimes \frac{\rho^{\rB} + r{\rho'}^{\rB}}{1+r}.
\end{eqnarray*}
By establishing the equivalence between the optimization problem (\ref{g}) when   Alice performs a full Bell state measurement and has access to a tomographically complete set of input states, and the optimization problem defining the generalized entanglement robustness, we can conclude that
\begin{equation}\label{GenTelEntEq}
\tau^*_{\textrm{gen}}(\{\sigma_{a\vert \omega_x}\}) = \epsilon_{\textrm{gen}}(\rho^{\rA\rB}).
\end{equation}

\subsubsection{Classical teleportation robustness}

For easier comparison let us restate the definition of the separable entanglement robustness, which is obtained from (\ref{EntRob}) with the constraint that $\rho_s$ is a separable state
\begin{eqnarray}\label{ClEntRob}
\epsilon_{\textrm{sep}}(\rho^{\rA\rB}) &=& \min_{r, \rho_s,\sigma_S} r\\ \nonumber
\textrm{s.t.}&\quad& \frac{\rho^{\rA\rB} + r \rho_s}{1+r} = \sigma_S\\ \nonumber
&\quad& \rho_s,\sigma_S \in \mathcal{S},
\end{eqnarray}
Let us, further, consider the classical teleportation robustness of a teleportation assemblage obtained when Alice applies a full Bell state measurement and uses a tomographically complete set of inputs, and denote it by $\tau^*_{\textrm{cl}}(\cdot)$. In order to reduce (\ref{cl}) to (\ref{ClEntRob}), it is useful to switch from the variables $\bar{M}_a$ to $p(a)$ and $\rho'_a$ which are related in the following way:
\begin{equation}\label{tilrho1}
dp(a)\left({\rho'}_a^{\rV\rB}\right)^{T_{\rV}} = \bar{M}_a^{\rV\rB}
\end{equation}
With this in place, members of the teleportation assemblage $\bar{\sigma}_{a|\omega_x}^{\rB}$ can be written in the form given in  ~(\ref{5}).

The simplification used in (\ref{k}) can again be used in exactly the same way, leading to 
\begin{multline} 
\textrm{tr}_{\rV}\left[\left(\frac{U_a^{\rV}\rho^{\rV\rB}{U_a^{\dagger}}^{\rV} + r d^2p(a){\rho'_a}^{\rV\rB}}{r+1}\right)^{T_{\rV}}\left(\omega_x^{\rV}\otimes \mathds{1}^{\rB}\right)\right]\\ \label{kcl}
=d\textrm{tr}_{\rV}[{M^*_a}^{\rV\rB}(\omega_x^{\rV} \otimes \mathds{1}^{\rB})].
\end{multline}
Since the set of input states is tomographically complete, (\ref{kcl}) implies 
\begin{equation*}
\frac{U_a^{\rV}\rho^{\rV\rB}{U_a^{\dagger}}^{\rV} + rd^2p(a){\rho'_a}^{\rV\rB}}{r+1}  = d\left({M^*_a}^{\rV\rB}\right)^{T_{\rV}}.
\end{equation*}
However, in this case $\rho'_a$ are separable (which is the consequence of  (\ref{tilrho1}) and the separability of  $\bar{M}_a^{\rV\rB}$). It is thus the case that the optimization (\ref{cl}) reduces to
\begin{subequations}\label{cl1}
\begin{align} 
&\tau^*_{\textrm{cl}}(\{\sigma_{a\vert \omega_x}\}) = \min_{r,\{\bar{M}_a, p(a)\rho'_a\}_a} r\\ \label{cl2}
\textrm{s.t.} &\quad   \frac{U_a^{\rV}\rho^{\rV\rB}{U_a^{\dagger}}^{\rV} + rd^2p(a){\rho'}_a^{\rV\rB}}{r+1}   = d\left({M^*_a}^{\rV\rB}\right)^{T_{\rV}},\\  \label{cl3}
&\quad  {M^*_a}^{\rV\rB} \geq 0, \quad {M^*_a}^{\rV\rB} \in \mathcal{S} \quad \forall a;  \\ \label{cl4}
&\quad  {\rho'_a}^{\rV\rB} \geq 0, \quad {\rho_a'}^{\rV\rB} \in \mathcal{S} \quad  \forall a;\\ \label{cl5}
&\quad   \sum_a p(a){\rho'_a}^{\rV\rB} = \frac{\mathds{1}^{\rV}}{d}\otimes \bar{\rho}^{\rB}; \\ \label{cl6}
&\quad   \sum_a M_a^{*\rV\rB} = \mathds{1}^{\rV}\otimes \frac{\rho^{\rB} + r\bar{\rho}^{\rB}}{1+r}.
\end{align}
\end{subequations}
In order to emphasize the resemblance with (\ref{ClEntRob}), let us rewrite (\ref{cl2}) in the following way
\begin{equation*}
 \frac{U_a^{\rV}\rho^{\rV\rB}{U_a^{\dagger}}^{\rV} + rd^2p(a){\rho'}_a^{\rV\rB}}{1+rd^2p_a} = \frac{1}{\tr{{M^*_a}^{\rV\rB}}}\left({M^*_a}^{\rV\rB}\right)^{T_{\rV}}.
\end{equation*}
Since all states that are mutually related by local unitary transformations have the same value for the separable entanglement robustness, the smallest $r$ satisfying the last equation for each $a$ is equal to $\epsilon_{\textrm{cl}}/d^2p(a)$. Analogously to the case of the generalized teleportation robustness, the optimal $r$ is obtained when $p(a) = 1/d^2$ and $\rho'_a = U_a\rho_0'U_a^{\dagger}$ for all values of $a$, and is equal to  $\epsilon_{\textrm{sep}}(\rho^{AB})$, which implies
\begin{equation}\label{cLTelEntEq}
\tau^*_{\textrm{cl}}(\{\sigma_{a\vert \omega_x}\}) = \epsilon_{\textrm{sep}}(\rho^{\rA\rB}).
\end{equation}

\subsubsection{Random teleportation robustness}

Finally, we consider the random teleportation robustness of a teleportation assemblage obtained when Alice applies a full Bell state measurement and has access to a tomographically complete set of inputs. Let us denote it accordingly by $\tau^*_{\textrm{r}}(\cdot)$.
We will compare it to the random entanglement robustness $\epsilon_{\textrm{r}}$:
\begin{eqnarray}\label{RanEntRob}
\epsilon_{\textrm{r}}(\rho^{\rA\rB}) &=& \min_{r, \sigma_S} \quad r\\ \nonumber
\textrm{s.t.}&\quad& \frac{\rho^{\rA\rB} +\mathds{1}/d^2}{1+r}  = \sigma_S\\ \nonumber
&\quad& \sigma_S \in \Sigma.
\end{eqnarray}
Recall that the definition of random teleportation robustness of a teleportation assemblage $\Tel$ is given in \eqref{r}. The first constraint of (\ref{r}) in the  case where Alice applies a Bell state measurement reads
\begin{eqnarray}\nonumber
&\quad& \qquad \qquad \frac{\sigma_{a\vert \omega_x}^{\rB} + rp(a)\mathds{1}^{\rB}/d}{1+r} \\ \nonumber &=& \textrm{tr}_{\rV\rA}\left[\left({\Phi^+}^{\rV\rA}\otimes \mathds{1}^{\rB}\right)\left( \omega_x^{\rV} \otimes \frac{U_a^{\rA}\rho^{\rA\rB}{U_a^{\dagger}}^{\rA} + rp(a)\mathds{1}^{\rA\rB}}{1+r} \right)\right]\\ \label{kolasin}
&=& \frac{1}{d}\textrm{tr}_{\rV}\left[\left( \frac{U_a^{\rV}\rho^{\rV\rB}{U_a^{\dagger}}^{\rV}+rp(a)\mathds{1}^{\rV\rB}}{1+r}\right)^{T_{\rV}}\left(\omega_x^{\rV}\otimes \mathds{1}^{\rB}\right)\right] \\
\nonumber &=& \textrm{tr}_{\rV} \left[{M^*_a}^{\rV\rB}(\omega_x^{\rV} \otimes \mathds{1}^{\rB})\right]
\end{eqnarray}
For a tomographically complete set of inputs this condition is satisfied if and only if
\begin{equation}\label{aerys}
\frac{U_a^{\rV}\rho^{\rV\rB}{U_a^{\dagger}}^{\rV}+rp(a)\mathds{1}^{\rA\rB}}{1+r} = d\left({M^*_a}^{\rV\rB}\right)^{T_{\rV}}
\end{equation}
Following this simplification, the optimization problem (\ref{r}) reduces to
\begin{subequations}\label{mojkovac}
\begin{align}\nonumber
\tau_{\textrm{r}}(\{\sigma_{a|\omega_x}\}) &= \min_{r, \{M^*_a\},\{p(a)\}} r\\\ \label{m2}
\textrm{s.t.} &\quad \frac{U_a^V\rho^{\rV\rB}{U_a^{\dagger}}^{\rV}+rp(a)\mathds{1}^{\rV\rB}}{1+r} = d\left({M^*_a}^{\rV\rB}\right)^{T_{\rV}} \\ \label{m3}
&\quad {M^*_a}^{\rV\rB} \geq 0, \quad {M^*_a}^{\rV\rB} \in \Sigma  \qquad \forall a, \\ \label{m4}
&\quad \sum_a{M_a^*}^{\rV\rB} = \mathds{1}\otimes \frac{\rho^{\rB} + r\frac{\mathds{1}^{\rB}}{d}}{1+r}.
\end{align}
\end{subequations}
For each value of $a$, (\ref{m2})  can be transformed in the following way
\begin{equation*}
\frac{U_a^{\rV}\rho^{\rV\rB}{U_a^{\dagger}}^{\rV}+rd^2p(a)\frac{\mathds{1}^{\rA\rB}}{d^2}}{1+rd^2p(a)} = \frac{1}{\tr{{M^*_a}^{\rV\rB}}}\left({M^*_a}^{\rV\rB}\right)^{T_{\rV}}.
\end{equation*}
Thus, the smallest $r$ satisfying (\ref{m2}) and (\ref{m3}) for each $a$ separately is equal to $\epsilon_{\textrm{r}}(U_a^V\rho^{\rV\rB}{U_a^{\dagger}}^{\rV})/d^2p(a) = \epsilon_{\textrm{r}}(\rho^{\rV\rB})/d^2p(a)$. Since there are $d^2$ different outcomes $a$, the smallest $r$ which can simultaneously satisfy (\ref{m2}) for all values of $a$ is equal to $\epsilon_{\textrm{r}}$. By summing \eqref{aerys} over $a$, we see that the last constraint from \eqref{r} is satisfied, which finally implies

\begin{equation*}
\tau_{\textrm{r}}^*(\sigma_{a\vert \omega_x}) = \epsilon_{\textrm{r}}(\rho^{\rA\rB}).
\end{equation*}

\subsection{Teleportation using a partial Bell state measurement}
Teleportation experiments where Alice performs a partial Bell state measurement using POVM $M_0^{\rV\rA} = {\Phi^+}^{\rV\rA}$, $M_1^{\rV\rA} = \sum_{i=1}^{d^2-1}U_i^{\rV}{\Phi^+}^{\rV\rA}{U_i^{\dagger}}^{\rV}$ and has access to a tomographically complete set of input states are also of particular interest.

Let us denote the random teleportation robustness of a teleportation assemblage obtained by performing such a measurement as $\tau'_{\textrm{r}}(\cdot)$. Taking into account that the set of input states is tomographically complete, $\tau'_{\textrm{r}}(\cdot)$  can be expressed as the solution to the following optimization problem
\begin{subequations}\label{gusinje}
\begin{align}\nonumber
\tau'_{\textrm{r}}(\{\sigma_{a|\omega_x}\}) &= \min_{r, \{M^*_a\},\{p(a)\}} r\\\ \label{a2}
\textrm{s.t.} &\quad \frac{\rho^{\rV\rB}+rp(0)\mathds{1}^{\rV\rB}}{1+r} = d\left({M^*_0}^{\rV\rB}\right)^{T_{\rV}}, \\ \label{a3}
&\quad \frac{\sum_{i=1}^{d^2-1}U_i^{\rV}\rho^{
\rV\rB}{U_i^{\dagger}}^{\rV}+rp(1)\mathds{1}^{\rV\rB}}{1+r} \nonumber \\ &\hspace{2.5cm}= d\left({M^*_1}^{\rV\rB}\right)^{T_{\rV}}, \\ \label{a4}
&\quad {M^*_a}^{\rV\rB} \geq 0, \quad {M^*_a}^{\rV\rB} \in \mathcal{S}  \qquad \forall a, \\ \label{a5}
&\quad \sum_a{M_a^*}^{\rV\rB} = \mathds{1}\otimes \frac{\rho^{\rB} + r\frac{\mathds{1}^{\rB}}{d}}{1+r}.
\end{align}
\end{subequations}
Note that the constraint (\ref{a3}), based on (\ref{a2}) and satisfying (\ref{a5}) can be reduced to
\begin{equation*}
{M^*_1}^{\rV\rB} = \sum_{i=1}^{d^2-1}{M^*_0}^{\rV\rB} + \frac{\left(p(1)-p(0)(d^2-1)\right)\mathds{1}^{\rV\rB}}{d(1+r)},
\end{equation*}
which is separable whenever ${M^*_0}^{\rV\rB}$ is separable\footnote{This is expected since constraint (\ref{a3}) corresponds to the member of teleportation assemblage which is obtained by using separable measurement $M_1^{\rV\rA}$.}. This means that every $r$ satisfying (\ref{a2}) also satisfies (\ref{a3}), which in turn implies that $\tau'_{\textrm{r}}(\{\sigma_{a|\omega_x}\})$ is equal to the smallest $r$ satisfying (\ref{a2}) and (\ref{a4}). Following the equivalence of (\ref{gusinje}) and (\ref{RanEntRob}), the smallest such $r$ is equal to $\epsilon_{\textrm{r}}(\rho^{\rA\rB})/d^2p(0))$. The optimal mixing assemblage is the trivial one $\{\mathds{1}^{\rB}/d, 0\}$ leading to

\begin{equation}
\tau'_{\textrm{r}}(\{\sigma_{a\vert \omega_x}\}) = \frac{\epsilon_{\textrm{r}}(\rho^{\rA\rB})}{d^2},
\end{equation}
Note that in \citep{CSS} a different bound was obtained, namely that $\tau'_{\textrm{r}}(\{\sigma_{a\vert \omega_x}\}) = \frac{2\epsilon_{\textrm{r}}(\rho^{\rA\rB})}{d^2}$. This is due to the different definition used for the random teleportation robustness. There, as noted earlier, the mixing teleportation assemblage had the form $\frac{\mathds{1}}{|o|d}$, which automatically fixed $p(0)$ to be equal to $1/2$. 

\section{Teleportation weight and best separable approximation} \label{twbsm}

In this appendix we first show that the teleportation weight of the teleportation assemblage $\{\sigma_{a|\omega_x}^{\rB}\}_{a,x}$  puts a lower bound on the best separable approximation of the shared state $\rho^{\rA\rB}$. 
First, let us observe that for the state $\rho^{\rA\rB}$ and its best separable approximation $\epsilon_\BSA(\rho^{\rA\rB})$ there exist a corresponding quantum state $\tilde{\rho}^{\rA\rB}$ and separable state $\bar{\rho}^{\rA\rB}$ such that
\begin{equation*}
\rho^{\rA\rB} = \epsilon_\BSA(\rho^{\rA\rB})\tilde{\rho}^{\rA\rB} + (1-\epsilon_\BSA(\rho^{\rA\rB}))\bar{\rho}^{\rA\rB}
\end{equation*}
By tensoring $\rho^{\rA\rB}$ with the state $\omega_x^{\rV}$ and applying a joint measurement $M_a^{\rV\rA}$, this implies
\begin{multline*}
\tr_{\rV\rA}\left[\left(M_a^{\rV\rA}\otimes \mathds{1}^{\rB}\right)\left(\omega_x^{\rV}\otimes \rho^{\rA\rB}\right)\right] = \\ \epsilon_\BSA(\rho^{\rA\rB})\tr_{\rV\rA}\left[\left(M_a^{\rV\rA}\otimes \mathds{1}^{\rB}\right)\left(\omega_x^{\rV}\otimes\tilde{\rho}^{\rA\rB}\right)\right] \\ + (1-\epsilon_\BSA(\rho^{\rA\rB}))\tr_{\rV\rA}\left[\left(M_a^{\rV\rA}\otimes \mathds{1}^{\rB}\right)\left(\omega_x^{\rV}\otimes\bar{\rho}^{\rA\rB}\right)\right],
\end{multline*}
i.e.
\begin{multline}\label{tak}
\sigma_{a\vert \omega_x}^\rB = \textrm{tr}_\rV\Big[\Big(\epsilon_\BSA(\rho^{\rA\rB})\tilde{M}_{a}^{\rV\rB} \\
+ \left(1-\epsilon_\BSA(\rho^{\rA\rB})\right)\bar{M}_{a}^{\rV\rB}\Big)\omega_x^\rV \otimes \mathds{1}^\rB\Big],
\end{multline}
where 
\begin{align}\label{kak}\begin{split}
\tilde{M}_{a}^{\rV\rB} &= \tr_\rA\left[\left(M_a^{\rV\rA}\otimes \mathds{1}^{\rB}\right)\left(\mathds{1}^{\rV}\otimes\tilde{\rho}^{\rA\rB}\right)\right],\\
 \bar{M}_{a}^{\rV\rB} &= \tr_\rA\left[\left(M_a^{\rV\rA}\otimes \mathds{1}^{\rB}\right)\left(\mathds{1}^{\rV}\otimes\bar{\rho}^{\rA\rB}\right)\right],\end{split}
 \end{align}
for all $a$ and $x$, \eqref{tak} is equivalent to the first constraint from the optimization problem \eqref{tw}. Moreover, the operators $\tilde{M}_{a}^{\rV\rB}$ and $\bar{M}_{a}^{\rV\rB}$ defined in \eqref{kak} satisfy all the other constraints from \eqref{tw}. Thus, the teleportation weight of the teleportation assemblage $\{\sigma_{a|\omega_x}\}_{a,x}$ can only be smaller than the best separable approximation of the shared state $\rho^{\rA\rB}$, i.e.
\begin{equation*}
\textrm{TW}(\{\sigma_{a|\omega_x}^{\rB}\}_{a,x}) \leq \epsilon_\BSA(\rho^{\rA\rB}).
\end{equation*}
Now we show that if the teleportation assemblage $\{\sigma_{a|\omega_x}^{\rB}\}_{a,x}$ is obtained by applying a Bell state measurement on Alice's share of the state $\rho^{\rA\rB}$ and states from a tomographically complete set $\{\omega_x\}_x$, its teleportation weight is equal to the best separable approximation of the state $\rho^{\rA\rB}$. In such a scenario the first constraint from the optimization problem \eqref{tw} can be rewritten in the following way
\begin{align*}
\begin{split}
\sigma_{a|\omega_x}^{\rB} &= \tr_{\rV\rA}\left[\left(U_a^{\rA}{\Phi^+}^{\rV\rA}{U_a^{\dagger}}^{\rA}\otimes \mathds{1}^{\rB}\right)\left(\omega_x^{\rV}\otimes \rho^{\rA\rB}\right)\right]\\
&= p\tr_{\rV\rA}\left[\left({\Phi^+}^{\rV\rA}\otimes \mathds{1}^{\rB}\right)\left(\omega_x^{\rV}\otimes d^2\tilde{p}(a)\tilde{\rho}_a^{\rA\rB}\right)\right] + \\ &+ (1-p)\tr_{\rV\rA}\left[\left({\Phi^+}^{\rV\rA}\otimes \mathds{1}^{\rB}\right)\left(\omega_x^{\rV}\otimes d^2\bar{p}(a)\bar{\rho}_a^{\rA\rB}\right)\right].
\end{split}
\end{align*}
The constraints on $\tilde{M}_a$ and $\bar{M}_a$ impose that the states $\tilde{\rho}_a$ could be any quantum states, while the states $\bar{\rho}_a$ are separable. Furthermore, using identity \eqref{MaxEntIdn} the last equation reduces to
\begin{multline}\label{twapp}
\frac{1}{d}\tr_{\rV}\left[\left(U_a^{\rV}\rho^{\rV\rB}{U_a^{\dagger}}^{\rV}\right)^{T_{\rV}}(\omega_x^{\rV}\otimes \mathds{1}^{\rB})\right] \\ = 
\frac{1}{d}\tr_{\rV}\left[\left(pd^2\tilde{p}(a)\tilde{\rho}_a^{\rV\rB} + (1-p)d^2\bar{p}(a)\bar{\rho}_a^{\rV\rB}\right)^{T_{\rV}}(\omega_x^{\rV}\otimes \mathds{1}^{\rB})\right].
\end{multline}
For tomographically complete set of inputs $\{\omega_x\}_x$ this equation is satisfied if and only if
\begin{equation}\label{twapp1}
U_a^{\rV}\rho^{\rV\rB}{U_a^{\dagger}}^{\rV} = d^2(p\tilde{p}(a)\tilde{\rho}_a^{\rV\rB} + (1-p)\bar{p}(a)\bar{\rho}_a^{\rV\rB}).
\end{equation}
Together with the constraints on the states $\tilde{p}_a$ and $\bar{p}_a$, we see that the optimization problem of finding the teleportation weight of a teleportation  assemblage obtained by using a Bell state measurement and tomographically complete set of inputs can be reduced to
\begin{align}\label{twapp2}\begin{split}
\textrm{TW}(\{&\sigma_{a\vert \omega_x}^\rB\}) = \min_
{p, \tilde{p}(a),\tilde{\rho}_a, \bar{p}(a), \bar{\rho}_a} p\\
  \textrm{s.t.} \quad &U_a^{\rV}\rho^{\rV\rB}{U_a^{\dagger}}^{\rV} = d^2(p\tilde{p}(a)\tilde{\rho}_a^{\rV\rB} + (1-p)\bar{p}(a)\bar{\rho}_a^{\rV\rB})\\
  & \sum_a\tilde{p}(a) = 1, \qquad \sum_a\bar{p}_a = 1\\
  & \bar{p}_a \in \mathcal{S}
\end{split}
\end{align}
For every $a$ the minimal $p$ satisfying \eqref{twapp1} is similar to the constraint appearing in the expression for the best separable approximation of the state $U_a^{\rV}\rho^{\rV\rB}{U_a^{\dagger\rV}}$. The difference is that in \eqref{twapp1} the states $d^2\tilde{p}(a)\tilde{\rho}_a^{\rV\rB}$ and $d^2\bar{p}(a)\bar{\rho}_a^{\rV\rB}$ need not be normalized. If for some $a = a'$, $\tr(d^2\tilde{p}(a')\tilde{\rho}_{a'}^{\rV\rB})$ is bigger than $1$, the minimal $p$ satisfying \eqref{twapp1} would be smaller than the best separable approximation of $U_{a'}^{\rV}\rho^{\rV\rB}{U_{a'}^{\dagger\rV}}$. But because $\sum_ad^2\tilde{p}(a)\tilde{\rho}_{a}^{\rV\rB} = d^2$ and there are $d^2$ different values of $a$, it means that for some other $a = a''$ we will have $\tr(d^2\tilde{p}(a'')\tilde{\rho}_{a''}^{\rV\rB}) < 1$, which would make the smallest $p$ satisfying \eqref{twapp1} strictly larger than the best separable approximation of $U_{a''}^{\rV}\rho^{\rV\rB}{U_{a''}^{\dagger\rV}}$. Since all the states related by local unitary transformations have the same best separable approximation, the optimal $p$ satisfying all $d^2$ different constraints contained in \eqref{twapp1} must be equal to $\epsilon_\BSA(\rho^{\rV\rB})$.

\end{appendix}


\begin{thebibliography}{30}

\bibitem{teleportation}C. H. Bennett, G. Brassard, C. Cr\'epeau, R. Jozsa, A. Peres, and W. K. Wootters, \emph{Teleporting an unknown quantum state via dual classical and Einstein-Podolsky-Rosen channels}, Phys. Rev. Lett. {\bf70}, 1895 (1993).

\bibitem{Gisin} N. Gisin, G. Ribordy, W. Titel, and H. Zbinden, \emph{Quantum cryptography}, Rev. Mod. Phys. {\bf74}, 145 (2002)

\bibitem{repeaters} H.-J. Briegel, W. D\"{u}r, J. I. Cirac, and P. Zoller, \emph{Quantum repeaters: the role of imperfect local operations in quantum communication}, Phys. Rev. Lett. {\bf81}, 5932 (1998)

\bibitem{Gottesman} D. Gottesman and I. L. Chuang, \emph{Demonstrating the viability of universal quantum computation using teleportation and single-qubit operations}, Nature {\bf402}, 390 (1999)

\bibitem{Raussendorf} R. Raussendorf and H.-J. Briegel, \emph{A one-way quantum computer}, Phys. Rev. Lett. {\bf86}, 5188 (2001) 

\bibitem{Sandu} S. Popescu, \emph{Bell’s inequalities versus teleportation: What is nonlocality?}, Phys. Rev. Lett, {\bf72}, 797 (1994)

\bibitem{Horodecki99}M. Horodecki, P. Horodecki, and R. Horodecki, \emph{General teleportation channel, singlet fraction, and quasidistillation}, Phys. Rev. A {\bf60}, 1888 (1999).

\bibitem{bound entanglement} M. Horodecki, P. Horodecki, and R. Horodecki, \emph{Mixed-State Entanglement and Distillation: Is there a “Bound” Entanglement in Nature?}, Phys. Rev. Lett. 80, 5239 (1998).

\bibitem{CSS} D. Cavalcanti, P. Skrzypczyk, and  I \v{S}upi\'{c} \emph{All entangled states can demonstrate non-classical teleportation},  Phys. Rev. Lett. {\bf119}, 110501 (2017) 

\bibitem{Rome} G. Carvacho, F. Andreoli, L. Santodonato, M. Bentivegna, V. D'Ambrosio, P. Skrzypczyk, I. \v{S}upi\'{c}, D. Cavalcanti, F. Sciarrino, \textit{Experimental study of nonclassical teleportation beyond average fidelity}, arXiv preprint 1802.10056 [quant-ph] (2018)







\bibitem{HobanSainz} M. J. Hoban and A. B. Sainz, \emph{A channel-based framework for steering, non-locality and beyond}, arXiv preprint 1708.00750 [quant-ph]

\bibitem{Doherty} A. C. Doherty, P. A. Parrilo, and F. M. Spedalieri, \textit{Distinguishing separable and entangled states}, Phys. Rev. Lett. \textbf{88}(18), 187904 (2002); A. C. Doherty, P. A. Parrilo, and F. M. Spedalieri, \textit{Complete family of separability criteria}, Phys. Rev. A \textbf{69}(2), 022308 (2004)

\bibitem{SDP} S. Boyd and L. Vandenberghe, \textit{Convex optimization}, Cambridge University Press (2004)

\bibitem{VW} G. Vidal and R. Werner, \textit{A computable measure of entanglement}, Phys. Rev. A \textbf{65} 032314 (2002)


\bibitem{Vidal} G. Vidal and R. Tarrach, \textit{Robustness of entanglement},  	Phys.Rev. A \textbf{59} 141-155 (1999)

\bibitem{Steiner} M. Steiner, \textit{Generalized robustness of entanglement}, Phys. Rev. A \textbf{67} 054305 (2003)

\bibitem{SCS}  I.\v{S}upi\'{c}, P. Skrzypczyk,   and D. Cavalcanti, \textit{Measurement-device-independent entanglement and randomness estimation in quantum networks} Phys. Rev. A
\textbf{95}, 042340 (2017).

\bibitem{Rosset} D. Rosset, A. Martin, E. Verbanis, C. C. W. Lim, R. Thew, \textit{Practical measurement-device-independent entanglement quantification }, arXiv preprint 1709.03090 [quant-ph] (2017)

\bibitem{BSA}M. Lewenstein and A. Sanpera, \emph{Separability and Entanglement of Composite Quantum Systems,} Phys. Rev. Lett. 80, 2261 (1998).

\bibitem{EPR2} A. C. Elitzur, S. Popescu, and D. Rohrlich, Phys. Lett. A162, 25 (1992).

\bibitem{St} P. Skrzypczyk, M. Navascu\'{e}s, D. Cavalcanti, \textit{Quantifying Einstein-Podolsky-Rosen steering}, Phys. Rev. Lett. \textbf{12}, 180404 (2014)

\bibitem{code} Notebook available at  \url{https://git.io/fN3Qh}

\bibitem{HorodeckiState} P. Horodecki \emph{Separability criterion and inseparable mixed states with positive partial transposition} Phys. Lett. A, {\bf232}:333, 1997

\bibitem{UPB} C. H. Bennet, D. P. DiVincenzo, T. Mor, P. W. Shor, J. A Smolin, B. Terhal, \emph{Unextendible product bases and bound entanglement},  Phys. Rev. Lett. {\bf82}, 5385 (1999)



\end{thebibliography}
\end{document}